\documentclass[preprint,12pt]{elsarticle}

\usepackage{lineno,hyperref}
\usepackage{bm}
\usepackage{amsfonts}
\usepackage{amssymb}
\usepackage{amsmath}
\usepackage{isomath}

\usepackage{scalerel,stackengine}
\stackMath
\newcommand\reallywidehat[1]{%
\savestack{\tmpbox}{\stretchto{%
  \scaleto{%
    \scalerel*[\widthof{\ensuremath{#1}}]{\kern-.6pt\bigwedge\kern-.6pt}%
    {\rule[-\textheight/2]{1ex}{\textheight}}
  }{\textheight}%
}{0.5ex}}%
\stackon[1pt]{#1}{\tmpbox}%
}

\usepackage{stmaryrd} 
\usepackage[usenames]{color}
\usepackage{epsfig}
\usepackage{graphicx}
\usepackage{psfrag}

\graphicspath{{Figures/}}
\usepackage{float}
\floatstyle{boxed}
\newfloat{algorithm}{htb}{loa}
\floatname{algorithm}{Algorithm}
\usepackage{algorithm,algpseudocode}
\usepackage{caption}
\usepackage{subcaption}
\usepackage{color}
\usepackage{xcolor}
\usepackage{multirow}
\usepackage{natbib}
\modulolinenumbers[5]

\usepackage[left = 2cm, right=2.5cm, top=2cm, bottom =2cm]{geometry}

\begin{document}
\begin{frontmatter}







\title{ \textcolor{black}{Learning Solutions of Thermodynamics-Based Nonlinear Constitutive Material Models using Physics-Informed Neural Networks}}

\author{ Shahed Rezaei$^{1,*}$, Ahmad Moeineddin$^{2}$, Ali Harandi$^{3}$ }
\address{$^1$ACCESS e.V., Intzestr. 5, D-52072 Aachen, Germany}
\address{$^2$Institute for Structural Analysis, Technical University of Dresden, \\Georg-Schumann-Str. 7, D-01187 Dresden, Germany}
\address{$^1$Institute of Applied Mechanics, \\ RWTH Aachen University, Mies-van-der-Rohe-Str. 1, D-52074 Aachen, Germany}
\address{$^*$ corresponding author: s.rezaei@access-technology.de}

\begin{abstract}
We applied physics-informed neural networks to solve the constitutive relations for nonlinear, path-dependent material behavior. As a result, the trained network not only satisfies all thermodynamic constraints but also instantly provides information about the current material state (i.e., free energy, stress, and the evolution of internal variables) under any given loading scenario without requiring initial data. One advantage of this work is that it bypasses the repetitive Newton iterations needed to solve nonlinear equations in complex material models. Additionally, strategies are provided to reduce the required order of derivative for obtaining the tangent operator.
The trained model can be directly used in any finite element package (or other numerical methods) as a user-defined material model. However, challenges remain in the proper definition of collocation points and in integrating several non-equality constraints that become active or non-active simultaneously. We tested this methodology on rate-independent processes such as the classical von Mises plasticity model with a nonlinear hardening law, as well as local damage models for interface cracking behavior with a nonlinear softening law.
In order to demonstrate the applicability of the methodology in handling complex path-dependency in a three-dimensional (3D) scenario, we tested the approach using the equations governing a damage model for a three-dimensional interface model. Such models are frequently employed for intergranular fracture at grain boundaries. We have observed a perfect agreement between the results obtained through the proposed methodology and those obtained using the classical approach. Furthermore, the proposed approach requires significantly less effort in terms of implementation and computing time compared to the traditional methods.
Finally, we discuss the potential and remaining challenges for future developments of this new approach.

\end{abstract} 

\end{frontmatter}

\section{Introduction} 

Implementing and evaluating nonlinear material models is a challenging yet essential task for achieving reliable predictions in many applications. 
For example, advanced models may be needed to represent the complex plastic behavior of metals, soils, and polymers, or to account for the softening of materials as damage and microcracks evolve \cite{ROTERS2019420, CHOO20181, BREPOLS201764}.
Typically, modelers use a thermodynamic framework to derive these models, which helps them to set up all the necessary equations and evolution laws for the internal state of the materials. Once this is done, complex codes must be programmed in a specific programming language, which can be a difficult and time-consuming process.
To solve material constitutive equations, the iterative Newton method is often used, which can be highly effective. However, this method can also become computationally expensive, as it requires repeating the algorithm for each integration point and nonlinear iteration within the finite element (FE) framework. 
In summary, developing and implementing nonlinear material models require significant expertise and effort, and access to appropriate computational resources. 
In addition to the previous point, material behavior is strongly influenced by lower scales. Multiscale material modeling involves solving a boundary value problem at the microscale. Once the microscale solution is obtained, the homogenized results are transferred to the structural problem at the macroscale \cite{MATOUS2017192}. However, the transformation of information between scales is a bottleneck in multiscale analysis, as it involves computationally-intensive tasks that can be time-consuming.

A recent trend in computational material mechanics is to use deep learning (DL) methods such as neural networks (NN) to bypass expensive calculations. The concept involves training a NN on a sufficient amount of data to predict material properties \cite{WEBER2022115384, Tipu_2022, met12030402} or using the trained NN as a surrogate model for a specific material system \cite{Bock2019, Fritzen2019}. See for example investigations by \citet{ALI2019205}, where authors coupled a NN model with a rate-dependant crystal plasticity FE method formulation and showed the predictive capability of the proposed model for non-proportional loading paths. The success of this approach relies heavily on the quality and quantity of available data as well as the machine learning optimizer used. \citet{HUANG2020113008} developed a machine learning based material modelling framework where the accumulated absolute strain is proposed to be the history variable of the plasticity model in order to account for the loading history. \citet{Frankel_2020} predicted the evolution of the stress field given the initial microstructure and external loading using convolutional NN. For more information, readers can refer to \cite{LIU2021109152, Dornheim2023, Ray2023}, as well as research and review papers in \cite{Peng2021, mianroodi2022lossless, Bastek2022, Henkes2022}. 

In the previous approach, the predictions of the neural network are solely based on the available data, which raises concerns about the reliability of the NN for cases outside the training range. However, recent studies have shown that it is possible to improve the NN's predictive accuracy for unseen situations by incorporating thermodynamic constraints into the training process \cite{Faisal2022}.
\citet{XU2021110072} introduced a NN architecture that outputs the Cholesky factor of the tangent stiffness matrix to weakly imposes convexity on the strain energy function.
\citet{LIU20191138} developed a data-driven multiscale material modeling method that integrates principles of homogenization theory and essential physics with machine learning techniques. See also \cite{Dey2022} for further extension of the method to reproduce the creep behavior of fiber-reinforced materials.
\citet{HEIDER2020112875} discussed the application of the informed-graph-based neural networks for anisotropic elasto-plastic behavior and proposed remedies to generate a frame-invariant machine-learning constitutive model. \citet{VLASSIS2021113695} introduced a DL framework to train smoothed elastoplasticity models with interpretable components, such as yield surface, and plastic flow that evolve based on a set of network predictions and showed that their approach provides accurate predictions of cyclic stress paths. \citet{FLASCHEL2023115867} developed an approach for the unsupervised discovery of material laws belonging to an unknown class of constitutive behavior (such as viscoelasticity and elastoplasticity). The main idea is to obtain the Helmholtz free energy and the dissipation potential from the data on displacements and reaction forces. \citet{FUHG2023115930} discussed a modular elastoplasticity formulation where different components of the model can be chosen to be either a phenomenological or a data-driven model depending on the amount of available information.
\citet{Weber2021} improved NN training with physical constraints for hyperelastic materials and showed that the introduced enhancements lead to better learning behavior.
\citet{Kalina2023} presented a data-driven multiscale framework utilizing physics-constrained NN for finite strain hyperelasticity problems. The idea is to set the Helmholtz free energy density as output and train the NN by using a set of invariants as the input of the NN. As a result, several physical principles are fulfilled automatically. See also investigations by \cite{KLEIN2022104703} on polyconvex neural network constitutive models.

\citet{MASI2021104277} proposed the thermodynamics-based NN for constitutive modeling. The authors proposed to learn the free energy of the material as a function of strain and internal state variables. See also investigations by \cite{D0SM00488J} on learning effective strain energy of elastic cellular metamaterials. \citet{MASI2023105245} also extended their methodology to allow for identifying, from data and first principles, admissible sets of internal variables in complex materials.

In addition, to feed-forward NNs, other DL methods such as recurrent neural networks (RNN) are also suitable choices for path-dependent material behavior \cite{Dornheim2023, Mozaffar2019, ZHANG2020102732}.
\citet{He_2022} proposed a data-driven constitutive modeling approach with the consideration of thermodynamics principles for path-dependent materials.  
\citet{BONATTI2022104697} considered an elastoplastic example and proposed an RNN architecture that respects self-consistency and performs best when applied to long sequences of small increments.
\citet{Koeppe2022} proposed a physics-explaining approach, which interprets RNN. The authors investigated case studies on elastoplasticity and viscoelasticity constitutive models.
\citet{DANOUN2022104436} presented a hybrid physics-AI-based model to predict non-linear mechanical behaviors via RNN where specific thermodynamical constraints are considered during the training phase. See also developments by \cite{Zhang2022}, \cite{MAIA2023115934} and \cite{rosenkranz2023comparative}.
In the reviewed articles, one requires enough data to train the NN.

Following the idea of physics-informed neural networks (PINN) \cite{RAISSI2019}, one can employ the physical laws as loss functions to enhance the network outcome \cite{HAGHIGHAT2021, Faroughi2022}. Once the underlying physics of the problem is completely known, the PINN can be trained without any initial data and solely based on the given set of equations \cite{REZAEI2022PINN, SAMANIEGO2020112790, FUHG2022110839}.
\citet{WEI2023115826} established a consistent neural network approach to model the constitutive behavior of interfaces by integrating physical conditions such as positive energy dissipation as additional training constraints in the loss function.
\citet{tipireddy2019comparative} investigated the use of PINN for learning unknown dynamics or constitutive relations of a dynamical system. According to the authors' findings, the trained models exhibit greater accuracy when tasked with learning a constitutive relation rather than the entire dynamics. 
\citet{HAGHIGHAT2023105828} presented a physics-informed neural network formulation for constitutive modeling where inequality constraints of elastoplasticity theory are embedded in the PINN loss functions. In \cite{HAGHIGHAT2021113741}, the authors illustrated an extension for PINN in solid mechanics for the case with von Mises elastoplasticity. 
See also investigation by \cite{NIU2023105177} for the extension to finite strain elastoplasticity and \cite{HE2023103531} for a deep energy-based method for classical elastoplasticity. \citet{EGHBALIAN2023105472} presented a NN architecture that is enriched with the physics of elastoplasticity and show that embedding this aspect enhanced the extrapolation capability for loading regimes outside the training data. 


The reviewed works make significant contributions to two main categories. In the first group, NNs are trained on sufficient data obtained from numerical simulations or experimental measurements (see for example \cite{JANG2021102919, fernandez2020}). To improve predictive accuracy, physical constraints are incorporated into these methods \cite{zhang2020thermodynamic, DANOUN2022104436, ASHERI2023112186, EGHBALIAN2023105472}. In the second group, authors applied PINN techniques to study nonlinear material behavior for specific boundary value problems \cite{HAGHIGHAT2021113741, NIU2023105177} or for the characterization of constitutive models \cite{HAGHIGHAT2023105828}.

The main novel contribution in this work is the application of PINN and the ability to satisfy thermodynamic constraints and predict real-time material states for any arbitrary loading without requiring initial data represents.
It is important to note that in this work, a simple feed-forward neural network (FFNN) is utilized, which can be easily integrated into any finite element computation. Additionally, automatic differentiation is employed to calculate the tangent operator, further simplifying the implementation process.
\color{black}

\begin{figure}[H] 
  \centering
  \includegraphics[width=0.90\linewidth]{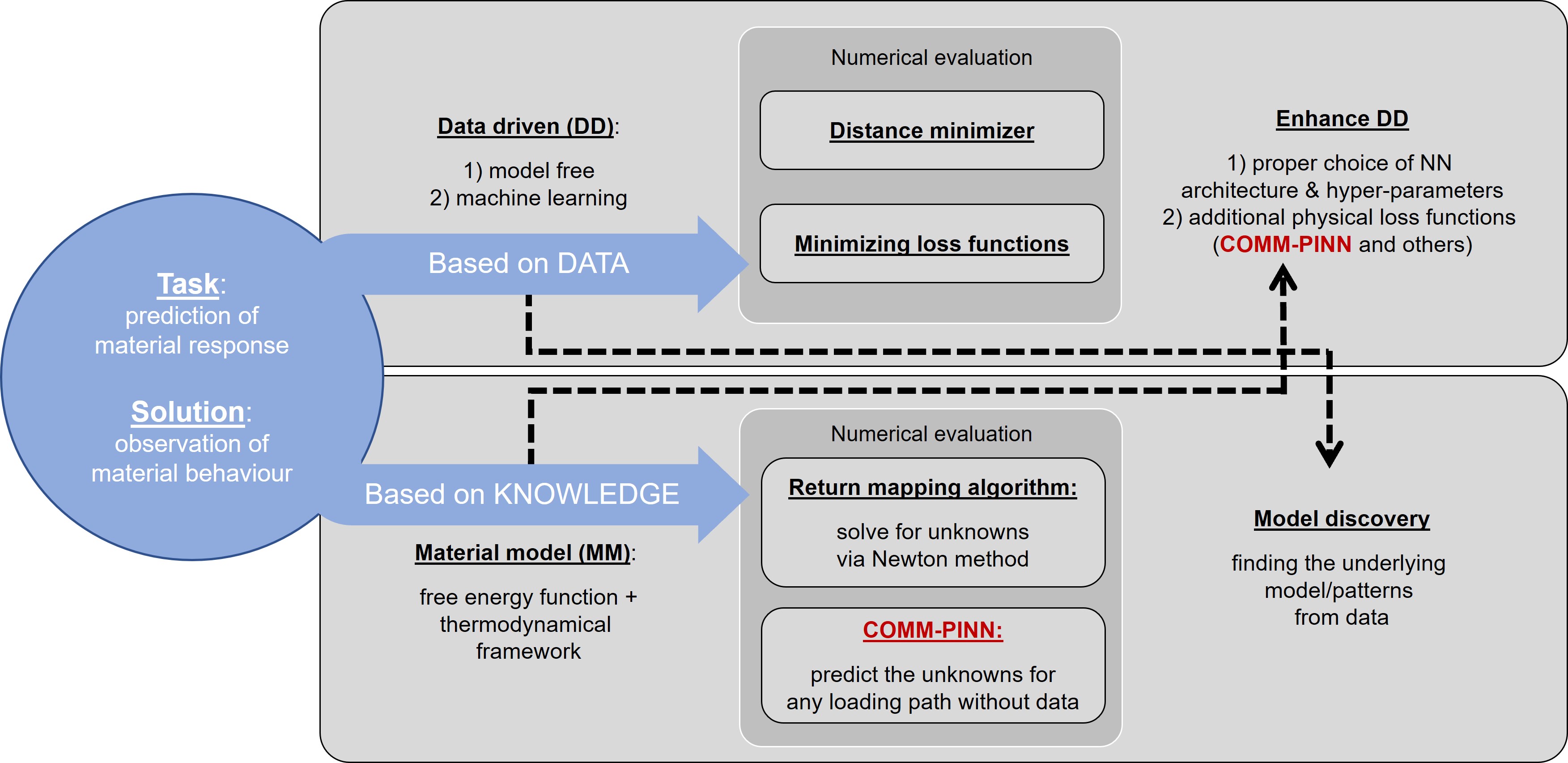}
  \caption{Summary of different approaches to predict of material behavior. The main idea of this proposed methodology (COMM-PINN) is to utilize neural networks to substitute the return mapping algorithm and enhance other data-driven approaches. References for each methodology are provided in the introduction part. }
  \label{fig:intro}
\end{figure}

To provide a clearer understanding for readers, Fig.~\ref{fig:intro} presents a summary of the available approaches for predicting complex and path-dependent material behavior under various loading conditions. Based on the available data from experiments or simulations, one can choose between data-driven (DD) approaches or physics-based methods. In physics-based modeling, the consistent equations of the model need to be numerically solved, and our proposed approach competes with the standard return-mapping algorithm. Notably, our approach does not rely on any data, distinguishing it from existing DD approaches.

In DD approaches, one can either employ a model-free approach \cite{EGGERSMANN201981} or train a deep learning model to fit the available data and construct a surrogate model \cite{Mozaffar2019}. In this context, the proposed COMM-PINN methodology can be integrated as an additional loss term within the neural network, aligning with similar approaches proposed by other authors as reviewed previously (e.g., \cite{WEI2023115826, EGHBALIAN2023105472, DANOUN2022104436, NIU2023105177, HAGHIGHAT2021113741}). 
\color{black}
 
Next, we will summarize the steps for deriving a consistent material model. We will also present case studies that involve elastoplastic behavior with a nonlinear isotropic hardening law, as well as a local damage model with a nonlinear softening law applied to model interface cracking. Finally, we will discuss the results and discuss possible future developments.

\section{Obtaining consistent constitutive relations and solving them with PINN}
The local form of dissipation inequality for the case of small deformation, pure mechanical case, and an isothermal process reads \cite{BREPOLS201764}

\begin{align}
\label{eq:CDI_main}
   -\dot{\psi}(\boldsymbol{\varepsilon}, \boldsymbol{\xi}_k) + \boldsymbol{\sigma}:\dot{\boldsymbol{\varepsilon}} \stackrel{!}{\ge} 0.
\end{align}
Here, $\psi=\psi(\boldsymbol{\varepsilon},\boldsymbol{\xi}_k)$ is the free energy of the material, $\boldsymbol{\sigma}$ is the stress tensor, $\boldsymbol{\xi}_k$ represent a set of internal (state) variables and $\boldsymbol{\varepsilon}$ is the strain tensor. The above inequality is simplified as
\begin{align}
\label{eq:CDI_r_main}
    (\underbrace{\boldsymbol{\sigma} - \partial_{\boldsymbol{\varepsilon}} \psi}_{=~\boldsymbol{0}})~\dot{\boldsymbol{\varepsilon}}
    -\Sigma_{k} \underbrace{ \partial_{\boldsymbol{\xi}_k} \psi}_{=~\boldsymbol{q}_k}~\dot{\boldsymbol{\xi}_k}
    \stackrel{!}{\ge} 0.
\end{align}
Next, the thermodynamic conjugate forces follow automatically as
\begin{align}
\label{eq:thermp_force}
   \boldsymbol{\sigma} = \partial_{\boldsymbol{\varepsilon}} \psi,~~~
    \boldsymbol{q}_k = \partial_{\boldsymbol{\xi}_k} \psi.
\end{align}
The remaining part of the dissipation inequality (i.e. $\mathcal{D}=-\Sigma_{k}  \boldsymbol{q}_k~\dot{\boldsymbol{\xi}_k} \stackrel{!}{\ge} 0$), still has to be satisfied for an arbitrary process.
To assure a positive dissipation rate and model the nonlinear material response, usually one introduces proper yield criterion $YLD(\boldsymbol{\xi}_k,...)$  as well as evolution equations $EVL(\boldsymbol{\xi}_k,...)$ for the state variables. In the following sections, we will investigate this part in more detail for two practical examples of plasticity and damage models.

\begin{figure}[H] 
  \centering
  \includegraphics[width=0.99\linewidth]{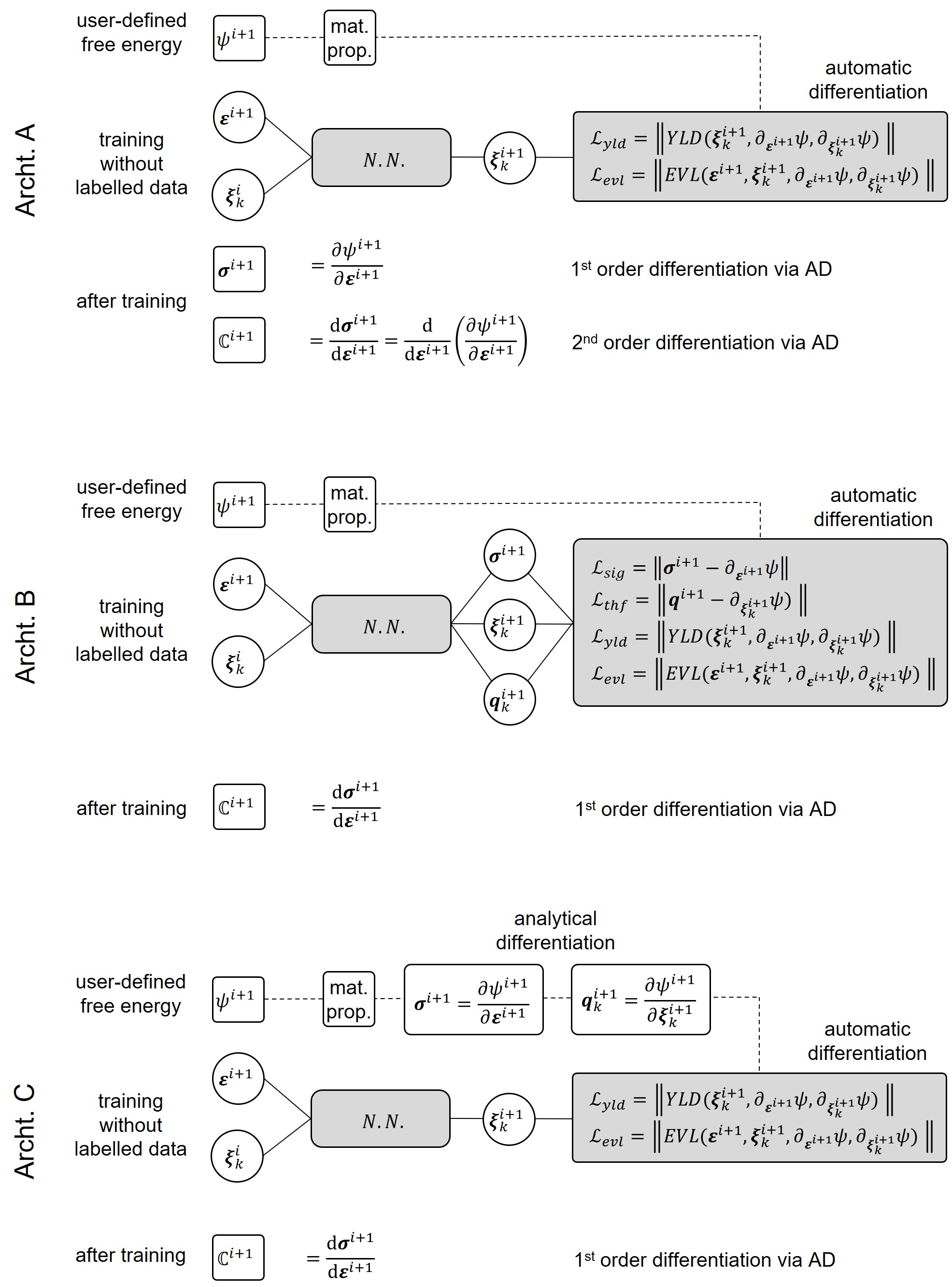}
  \caption{ Three suggestions as potential neural network architectures and loss functions for solving constitutive material equations via physics-informed neural networks (COMM-PINN).}
  \label{fig:summ}
\end{figure}

At this point, we have all the ingredients to set up a neural network for predicting nonlinear material behavior. In Fig.~\ref{fig:summ}, three architectures are proposed for solving equations in an arbitrary \underline{co}nstitutive \underline{m}aterial \underline{m}odel with \underline{p}hysics \underline{i}nformed \underline{n}eural \underline{n}etworks (COMM-PINN). 

Next, we will examine the pros and cons of each suggested network. In all cases, the user defines the free energy function and the required state variables. Furthermore, the inputs of the network are the current strain and the history of the material.

For case A, only the state variables are the outputs. All the thermodynamic conjugate forces ($\boldsymbol{\sigma}^{i+1}$ and $\boldsymbol{q}_k^{i+1}$) and the consistent tangent operator $\mathbb{C}^{i+1}=\text{d}\boldsymbol{\sigma}^{i+1}/\text{d}\boldsymbol{\varepsilon}^{i+1}$ are calculated via automatic differentiation (AD). One possible drawback is the requirement for second-order differentiation, for which proper activation functions must be used \cite{MASI2021104277}. Loss functions in this case are based on the yield function and evolution laws.

For case B, the thermodynamic conjugate forces are defined as an additional output, and their definitions are used as additional loss functions (see \cite{REZAEI2022108177} and references therein for breaking the order of derivative in PINN). One advantage of the second architecture is that we only require first-order derivatives to obtain the tangent operator. However, the network in this case might be denser to handle multiple outputs. In both of the above cases, the only requirement from the user is to define the free energy function, and all the thermodynamic forces, as well as updated state variables, are natural outcomes of the network. Finally, it is important to note that training the neural network, in this case, can be particularly challenging due to the entangled nature of the equations involved and the numerous loss terms that need to be considered.

For case C, we make training easier by taking advantage of the analytical differentiation of the free energy function and feeding the network with the analytical expression for the thermodynamic conjugate forces, such as the stress tensor. Obtaining such derivations can be a drawback for complicated energy functions, but it is usually straightforward to get such partial derivatives. The advantage is fewer loss terms, less entanglement of equations, and a first-order derivative for obtaining the tangent operator. An alternative approach is to utilize analytical differentiation to obtain the tangent operator, thereby avoiding the need for AD. While this may not be suitable for more complex material models, it presents an intriguing option worth considering.

\newpage
\section{A case study on von mises plasticity model} 
\subsection{Model derivation}
We limit ourselves to a 1D setting for small deformation. Nevertheless, the current investigations can be extended to higher dimensions and also case studies where more nonlinear behavior is expected. The strain variable $\varepsilon$ is defined as derivative of displacement $u$ with respect to the coordinate $x$ (i.e. $\varepsilon=\text{d}u/\text{d}x$). In the case of elastoplastic behavior, the strain variable is additively decomposed into an elastic and a plastic part as
\begin{equation}
\label{eq:eps}
   \varepsilon = \varepsilon_{e} + \varepsilon_{p}.
\end{equation}
The Helmholtz free energy $\psi$ is a contribution of the elastic part ($\psi_e$) and
the plastic part ($\psi_p$). Therefore, we write the following expressions
\begin{align}
\label{eq:tot_en}
   \psi(\varepsilon,\varepsilon_p,\xi_p) &= \psi_e(\varepsilon,\varepsilon_p)  + \psi_p(\xi_p), \\
   \label{eq:el_en}
   \psi_e \left( \varepsilon,\varepsilon_{p} \right) &= 
   \dfrac{1}{2} E (\varepsilon-\varepsilon_{p})^2, \\
   \psi_p \left( \xi_{p} \right) &= 
   h_1 \left( \xi_{p} + \dfrac{e^{-h_2\xi_{p}}-1}{h_2}  \right).
\end{align}
In the elastic part of the energy ($\psi_e$), we have the material Young's modulus $E$. Moreover, in the plastic part of the energy we introduce the parameter $\xi_{p}$ which represents the accumulative plastic strain. Only isotropic hardening is considered and the constants $h_1$ and $h_2$ are corresponding plasticity hardening parameters. Using the chain rule of differentiation and considering the additive elastoplastic split of the strain rate, the Clausius-Duhem inequality in its isothermal local form reads
\begin{align}
\label{eq:CDI_2}
    (\sigma - \partial_{\varepsilon} \psi)~\dot{\varepsilon}
    - \partial_{\epsilon_p} \psi~\dot{\varepsilon}_p
    - \partial_{\xi_p} \psi~\dot{\xi}_p
    \stackrel{!}{\ge} 0.
\end{align}
The above expression must hold for arbitrary thermomechanical processes. A common choice is to set the expressions in the bracket to zero. The state relations of the model as well as the thermodynamic conjugate forces follow automatically as
\begin{align}
\label{eq:stress}
   \sigma &= \partial_{\varepsilon} \psi = \partial_{\varepsilon} \psi_e  = 
   E (\varepsilon-\varepsilon_{p}),\\
   q_p &= \partial_{\xi_p} \psi = 
   \partial_{\xi_p} \psi_p                                = h_1~(1-e^{-h_2\xi_{p}}).
\end{align}
The remaining part of the dissipation inequality can now be written as
\begin{align}
\label{eq:CDI_3}
    \sigma~\dot{\varepsilon}_p
    - q_p~\dot{\xi}_p
    \stackrel{!}{\ge} 0.
\end{align}
The plasticity yield criterion (based on the von Mieses model) is written as
\begin{align}
\label{eq:pl_y}
   \phi_p &= |\sigma| - \left( \sigma_{y0} + {q}_p \right). 
\end{align}
Moreover, the following evolution laws for the plastic internal variables are used:
\begin{align}
\label{eq:pl_ev}
   \dot{\varepsilon}_{p} &= \dot{\lambda}_p~\dfrac{\partial \phi_p}{\partial \sigma} = \dot{\lambda}_p~\text{sgn}(\sigma), \\
\label{eq:pl_ev_2}
   \dot{\xi}_{p} &= -\dot{\lambda}_p~\dfrac{\partial \phi_p}{\partial q_p} = \dot{\lambda}_p.
\end{align}
Finally, the plastic loading/unloading conditions of the model are taken into account:
\begin{align}
\label{eq:pl_kkt}
   \dot{\lambda}_p\geq0,~~~\phi_p\leq0,~~~\dot{\lambda}_p~\phi_p=0.
\end{align}
By considering the relations \ref{eq:pl_ev}, \ref{eq:pl_ev_2} as well as constraints in \ref{eq:pl_kkt}, the nonequality \ref{eq:CDI_3} is satisfied.

\subsection{Implementation and algorithmic aspects for the plasticity model}
Algorithm~\ref{alg:plas} summarizes the return mapping algorithm commonly used to solve the plasticity equations. In order to solve for the unknowns, the equations for plasticity (or any other nonlinear material routines) must be linearized, which is accomplished using a local Newton solver at each integration point of the finite element method. Additionally, the user must derive equations for the tangent operator to be used in finite element programs. These tasks must be repeated for each global iteration step at the structural level and for each integration point in the domain, making them computationally expensive.

\begin{algorithm}[H]
\caption{Solving the elastoplastic governing equations at pseudo time $t^{i+1}$}
\label{alg:plas}
\begin{algorithmic}[1]
\item[\textbf{Input:}] material parameters, current strain: $\varepsilon^{i+1}$, history variables: $\varepsilon^i_p$, $\xi^i_p$
\item[\textbf{Output:}] current stress: $\sigma^{i+1}$, current history variables: $\varepsilon^{i+1}_p$, $\xi^{i+1}_p$, tangent: $C_p^{i+1}=\text{d} \sigma^{i+1} / \text{d} \varepsilon^{i+1}$
\State $k=1$
\State Set trial values: $\varepsilon^{i+1, k}_p=\varepsilon^{i}_p$, $\xi^{i+1, k}_p=\xi^i_p$
\State \textbf{loop} \textit{plasticity solver} \textbf{do} \\
\vspace{1mm}
$\qquad $ \textbf{if} $\phi^{tr}_p = |E(\varepsilon^{i+1}-\varepsilon^{i}_p)| - \left( \sigma_{y0} + {q}_p(\xi^{i}_p) \right) \le 0$ \\
\vspace{1mm}
$\qquad $ $\qquad $   $\varepsilon^{i+1}_p=\varepsilon^{i}_p$ and $\xi^{i+1}_p=\xi^{i}_p$ and $C^{i+1}= E$ \\ 
\vspace{1mm}
$\qquad $ \textbf{else} solve for $\varepsilon^{i+1,k+1}_p$ and $\xi^{i+1,k+1}_p = \xi^{i}_p+\Delta \lambda^{k+1}_p$ (see Eqs~\ref{eq:pl_y},\ref{eq:pl_ev})\\ 
\vspace{1mm}
$\qquad $  $\qquad  {r}^{(1)}_p = \varepsilon^{i+1,k+1}_p - \varepsilon^i_p - (\xi^{i+1,k+1}_p - \xi^{i}_p) \dfrac{\partial \phi_p}{\partial \sigma} \stackrel{!}{=} {0}$ \\
$\qquad $  $\qquad  r^{(2)}_p  =   \phi^{i+1}_p  = |\sigma^{i+1,k+1}|  - \left( \sigma_{y0} + {q}_p\left( \xi^{i+1}_p \right) \right) \stackrel{!}{=} 0$ \\
\vspace{1mm}
$\qquad$  $\qquad [\varepsilon^{i+1,k+1}_p~~~ \xi^{i+1,k+1}_p]^T = [\varepsilon^{i+1,k}_p~~~\xi^{i+1,k}_p]^T - \boldsymbol{K}_p^{-1} \boldsymbol{r}_p$ \\
$\qquad $ \textbf{end if}\\
\vspace{1mm}
$\qquad $ \textbf{if} $\left| \varepsilon^{i+1,k+1}_p-\varepsilon^{i+1,k}_p\right|>\text{tol} $ \textbf{OR} $\left| \xi^{i+1,k+1}_p-\xi^{i+1,k}_p\right|>\text{tol} $  \\
$\qquad $ $\qquad $   $k++$ \\
$\qquad $ $\qquad $   \textbf{CYCLE loop} \textit{plasticity} \\ 
$\qquad $ \textbf{else} \\
$\qquad $ $\qquad $   ${g}^{i+1}_p={g}^{i+1,k+1}_p$, $\xi^{i+1}_p=\xi^{i+1,k+1}_p$ \\ 
$\qquad $ $\qquad $   \textbf{EXIT loop} \textit{plasticity solver} \\
$\qquad $ \textbf{end if}
\vspace{1mm}
\State \textbf{end loop}
\State Compute stress $\sigma^{i+1}=E (\varepsilon^{i+1}-\varepsilon^{i+1}_p)$ see Eq.~\ref{eq:tra}
\State Compute tangent $C^{i+1}= E(1- \frac{E}{E + h_1 h_2~ \text{exp}(-h_2 \xi^{i+1}_p)})$
\end{algorithmic}
\end{algorithm}

The superindex $k$ in Algorithm~\ref{alg:plas} shows the $k-$th iteration in the internal loop \textit{plasticity-solver}. Furthermore, we used the notation $\Delta \lambda_p=\Delta t~\lambda_p$, where $\Delta t$ is the pseudo time step. Once the plasticity is active ($\phi_p > 0$) the plastic residuals (${r}^{(1)}_p$ and $r^{(2)}_p$) are linearized according to the Newton-Raphson method to find the solution of the unknowns ($\varepsilon^{i+1,k+1}_p$ and $\xi^{i+1,k+1}_p$). The matrix $\boldsymbol{K}_p = \partial \boldsymbol{r}_p / \partial \boldsymbol{U}_p$ includes the derivatives of the plastic residual vector $\boldsymbol{r}_p$ with respect to the unknowns ($\boldsymbol{U}_p=\{\varepsilon^{i+1}_p,\xi^{i+1}_p\}$). If the changes in the solution of the unknown internal variables are smaller than a certain tolerance ($\text{tol}=10^{-10}$), the obtained results are considered as the converged values and will be used to compute the stress value and the tangent operator. Finally, we are able to set up the global residuals at the finite element level \cite{BREPOLS201764, REZAEI2019325}.

\subsection{PINN to solve for elasto-plastic behavior with nonlinear isotropic hardening law} \label{sec:plas}
Based on the methodologies proposed by \citet{RAISSI2019}, the input for the network is the location of the collocation points. In the context of material modeling, the collocation points include the information on the current loading state (applied strain/gap) as well as admissible values for history variables (i.e. plastic strain, damage, etc.).
Here we propose to use architecture C according to Fig.~\ref{fig:summ}. Therefore, the output layer includes the current (updated) history variables (i.e. new plastic strain, damage, etc.). The other quantities such as stress, energy, and tangent operator are then calculated based on the known free energy equation of the material. In other words, we do not need any additional NN for computing those quantities.  

The structure of each neural network takes the standard form where it can be split into a single input layer, several possible hidden layers, and the output layer. Each layer is connected to the next layer for transferring information \cite{schmidhuber2015deep}. In every single layer, there is no connection among its neurons. Therefore, we represent the information bypass from the $l-1$ layer to $l$ via the vector $\bm{z}^l$. Every component of vector $\bm{z}^l$ is computed by 
\begin{equation}
\label{eq:NN_1}
    {z}^l_m = {a} (\sum_{n=1}^{N_l} w^l_{mn} {z}_n^{l-1} + b^l_{m} ),~~~l=1,\ldots,L. 
\end{equation}
In Eq.\,(\ref{eq:NN_1}), ${z}^{l-1}_n$, is the $n$-th neuron within the $l-1$-th layer. The component $w_{mn}$ shows the connection weight between the $n$-th neuron of the layer $l-1$ and the $m$-th neuron of the layer $l$. Every individual neuron in the $l$-th hidden layer owns a bias variable $b_m^l$. The number $N_I$ corresponds to the number of neurons in the $l$-th hidden layer. The total number of hidden layers is $L$. The letter $a$ stands for the activation function in each neuron. The activation function $a(\cdot)$ is usually a non-linear function. 
The proper choice of the activation function is problem dependent and shall be obtained based on hyperparameter studies \cite{REZAEI2022PINN}, \cite{Jagtaop22}. 

\begin{figure}[H] 
  \centering
  \includegraphics[width=0.99\linewidth]{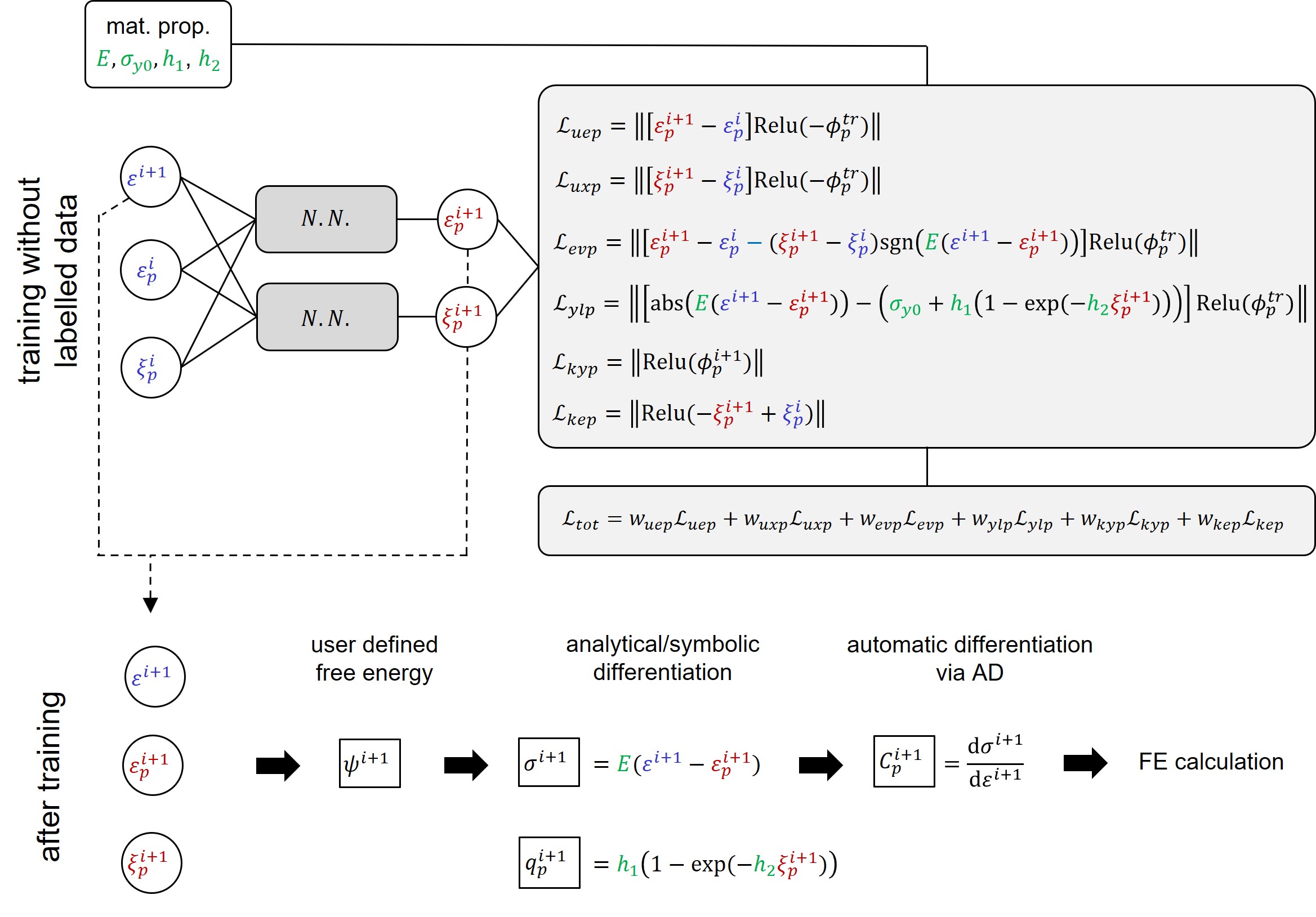}
  \caption{ Network architecture and loss functions for the COMM-PINNs applied to elasto-plastic material models with isotropic Voce type hardening law. }
  \label{fig:plas_pinn}
\end{figure}

For the case of plasticity, the input layer contains $\varepsilon^{i+1}$, $\varepsilon^{i}_p$ and $\xi^{i}_p$. We denote the input layer via vector $\bm{X}=\{\varepsilon^{i+1}, \varepsilon^{i}_p,\varepsilon^{i}_p\}$. The output layer is written as $\bm{Y}=\{\varepsilon^{i+1}_p, \xi^{i+1}_p\}$. We use separate fully connected FFNN for each output variable (see also Fig.~\ref{fig:plas_pinn}).
The trainable set of parameters of the network is represented by $\bm{\theta}=\{ \bm{W},\bm{b}\}$, which are the weights and biases of a neural network, respectively. Considering each neural network structure as $\mathcal{N}$, the outcome of the constitutive material modeling via neural network reads
\begin{align}
\label{eq:NN_pls}
    \varepsilon^{i+1}_p = \mathcal{N}_{\varepsilon_p} (\bm{X}; \bm{\theta}),~~~~~
    \xi^{i+1}_p      = \mathcal{N}_{\xi_p} (\bm{X}; \bm{\theta}).
\end{align}

The neural network outputs are functions of the trainable parameters and the training is done by minimizing physical loss functions. Next, we build the residuals and conditions for the introduced plasticity model in terms of the defined input and output layers. To do so, one requires to integrate the yield function as well as evolution equations for the internal variables into loss functions for the neural networks.
By denoting the summation of total loss terms for plasticity by $\mathcal{L}_{pt}$, and based on the Alg.~\ref{alg:plas}, one defines $\mathcal{L}_{pt}$ as
\begin{align}
\label{Totalloss_plas}
\mathcal{L}_{pt} &= 
\underbrace{w_{uep}\mathcal{L}_{uep} + w_{uxp}\mathcal{L}_{uxp}}_{\text{elastic response}} + \underbrace{w_{evp}\mathcal{L}_{evp}  + w_{ylp}\mathcal{L}_{ylp}}_{\text{plastic evolution}} + \underbrace{w_{kep}\mathcal{L}_{kep}  + w_{kyp}\mathcal{L}_{kyp}}_{\text{KKT conditions}}.
\end{align}
In Eq.\,(\ref{Totalloss_plas}), different loss terms cover all the possible loading, unloading, and reloading scenarios. The first two terms ($\mathcal{L}_{uep}$ and $\mathcal{L}_{uxp}$) guarantee that there is no evolution of plastic strain and accumulative plastic strain when the trial yield function ($\phi^{tr}_p$) is negative. Once $\phi^{tr}_p>0$, the main plastic residuals for evolution law as well as the current yield function become active which are denoted by $\mathcal{L}_{evp}$ and $\mathcal{L}_{ylp}$, respectively. Finally, to make sure that the Karush–Kuhn–Tucker conditions (KKT) conditions are always satisfied, we have the last two loss terms ($\mathcal{L}_{kep}$ and $\mathcal{L}_{kyp}$).
All these relevant loss terms are summarized in what follows
\begin{align}  
\label{loss_uep}
\mathcal{L}_{uep} &= \text{MSE}\left( (\varepsilon^{i+1}_p-\varepsilon^{i}_p)\text{Relu}(-\phi^{tr}_p) \right), \\
\label{loss_uxp}
\mathcal{L}_{uxp} &= \text{MSE}\left( (\xi^{i+1}_p-\xi^{i}_p)\text{Relu}(-\phi^{tr}_p) \right), \\
\label{loss_evp}
\mathcal{L}_{evp} &= \text{MSE}\left( (\varepsilon^{i+1}_p-\varepsilon^{i}_p-(\xi^{i+1}_p-\xi^{i}_p)\text{sgn}(E(\epsilon^{i+1}-\varepsilon^{i+1}_p)))\text{Relu}(\phi^{tr}_p) \right), \\
\label{loss_ylp}
\mathcal{L}_{ylp} &= \text{MSE}\left( \left( \text{abs}(E(\varepsilon^{i+1}-\varepsilon^{i+1}_p)) - ( \sigma_{y0} + h_1(1-\text{exp}(-h_2\xi^{i+1}_p))) \right)\text{Relu}(\phi^{tr}_p) \right), \\
\label{loss_kep}  
\mathcal{L}_{kep} &= \text{MSE}\left( \text{Relu}(\phi^{i+1}_p) \right), \\
\label{loss_kyp}
\mathcal{L}_{kyp} &= \text{MSE} \left( \text{Relu}(-\xi^{i+1}_p+\xi^{i}_p) \right).
\end{align}
In the above relation, $\phi^{tr}_p$ is the so-called trial yield function which is evaluated by means of the (known) quantities at the input layer (see also Alg.~\ref{alg:plas}):
\begin{align}
\label{trial_pl}
\phi^{tr}_p = \text{abs}(E(\varepsilon^{i+1}-\varepsilon^{i}_p)) - \left( \sigma_{y0} + h_1(1-\text{exp}(-h_2\xi^{i}_p)) \right).
\end{align}
For completeness, the expressions ``abs'' and ``sgn'' represent absolute value and sign function, respectively. The mean squared error is defined as
\begin{align}
\label{eq:MSE}
\text{MSE}( \bullet )_{type} &= \dfrac{1}{n}\sum_{i=1}^{n}(\bullet)^2, 
\end{align}
where $n$ is the number of observation points. The final loss term is minimized at every single collocation point. The mathematical optimization problem is written as  
\begin{align}
\label{minimize}
\bm{\theta}^* = \arg \min_{\bm{\theta}} \mathcal{L}_{pt}(\bm{X}; \bm{\theta}),
\end{align}
where $\bm{\theta}^*$ are the optimal trainable parameters (weights and biases) of the network. 
\\ \\
\textbf{Remark 1} In the above mentioned loss terms, the ``Relu'' function acts as a switch (if condition). In other words, we differentiate between elastic loading/unloading and the evolution of plastic variables via the sign of the $\phi^{tr}_p$. See also the same procedure in classical iterative solvers (i.e. Alg.~\ref{alg:plas}). One can also use other options such as ``sgn'' function for this purpose. Based on our studies, we realized that utilizing ``Relu'' functions is more beneficial.
\\ \\
In Fig.~\ref{fig:plas_pinn}, we summarized all loss terms and the idea behind constitutive material modeling via physics-informed neural networks. For the sake of clarity, all the input variables are denoted in blue color while the output (unknown) variables are presented in red color. Note that the material properties for the time being are kept constant and represented by green color.

After the training is completed, the network predicts the unknown internal variables (i.e. $\varepsilon^{i+1}_p$ and $\xi^{i+1}_p$ for the case of elastoplasticity). Having the updated internal variables as well as the current strain ($\varepsilon^{i+1}$), one can construct the free energy function ($\psi^{i+1}$) as well as the stress value ($\sigma^{i+1}$). Note that the functionality of stress can be obtained by analytical differentiation of the energy function with respect to the strain which is according to the thermodynamical framework \cite{MASI2021104277, Kalina2023}, see also architecture C in Fig.~\ref{fig:summ}. The final important parameter for the FE calculations is the tangent operator which is defined as the full derivative of the current stress value with respect to the current strain value. This parameter is obtained via automatic differentiation of the trained network. The main difference of the current work compared to other studies (especially those in \cite{MASI2021104277}), is the fact that here we do not provide any data to the FFNN.

\noindent
\textbf{Remark 2} Although the material properties are kept constant in this study, one can also include them as additional input parameters. As a result, the NN learns the solution to the elastoplastic problem for vast range of given material properties. See investigations by \cite{REZAEI2022PINN, harandi2023mixed} in direction of using transfer learning to include the influence of material properties.  
\\ \\
\textbf{Remark 3} In the above loss functions, one can add the contribution of the data in such form: $\mathcal{L}_{data} = \text{MSE}\left(\varepsilon^{i+1}_p-\varepsilon^{i+1}_{alg}\right) + \text{MSE}\left(\xi^{i+1}_p-\xi^{i+1}_{alg}\right)$, where subindex $alg$ represent the data coming from the return-mapping algorithm. We examine and compare such network designs in the result section.
\\ \\
\textbf{Remark 4} It is possible to derive the terms for the tangent operator via analytical derivative. The advantage is that one can insert the current values of strain and other state variables to find these quantities without running into the problem of vanishing gradients. The disadvantage is the possible difficulties in obtaining the analytical derivative. In the case of complicated differentiation, one can also use recent advances in parametric differentiation to avoid tedious derivations.

\section{A case study on damage model for interface cracking}
\subsection{Model derivation}
The gap value ${g}$ is utilized to formulate the interface energy and mechanics. The gap variable is related to the displacement jump $\langle \boldsymbol{u} \rangle$ or the distance between two opposite sides of the interface. Under mode I opening, we write the normal gap variable as
\begin{equation}
\label{eq:gap}
   {g} = \langle \boldsymbol{u} \rangle \cdot \boldsymbol{n},
\end{equation}
where $\boldsymbol{n}$ is the normal vector to the interface.
Damage at the interface is represented via a scalar parameter $d$ which is an internal variable and describes the loss of bonding forces. The (normal) traction parameter ${T}$ is related to the gap value ${g}$ and other internal variables such as damage $d$. Similar to the stress-strain behavior for bulk we have the traction-separation relation for the mechanics of a sharp interface.
The Helmholtz free energy $\psi$ at the interface is a contribution of the elastic part ($\psi_e$) and damage hardening ($\psi_d$) which lead in total to the following expression:
\begin{align}
    \label{eq:tot_en_d}
   \psi(g,d,\xi_d) &= \psi_e(g,d)  + \psi_d(\xi_d), \\
   \psi_e \left( \varepsilon,\varepsilon_{p} \right) &= 
   \label{eq:el_en_d}
   \dfrac{1}{2} f_{\text{d}}(d)~K~g^2, \\
   \label{eq:dm_en}
   \psi_d \left( \xi_{d} \right) &= 
   h_1 \left( \xi_{d} + \dfrac{e^{-h_2\xi_{d}}-1}{h_2}  \right),
\end{align}
Here, $\xi_{d}$ is damage hardening variables. Moreover, $h_1$ and $h_2$ are interface damage hardening parameters and control the softening behavior \cite{REZAEI2019325}. The parameter ${K}$ is the initial stiffness of the interface. Furthermore, $f_{\text{d}}(d)$ is a positive scalar value that varies between $1$ for an undamaged interface and a critical value close to $0$ for a completely damaged interface. A common choice for this function is as follows
\begin{equation}
\label{eq:fd}
f_{\text{d}}(d)=(1-d)^2.
\end{equation}

Applying the second law of thermodynamics in terms of the Clausius-Duhem inequality, consistent state relations of the model are derived. The conjugate forces are as follows 
\begin{align}
\label{eq:tra}
   {T} &= \partial_{g} \psi = \partial_{g} \psi_e = f_{\text{d}}(d) K~g,\\
    \label{eq:Y}
   Y   &= -\partial_{d} \psi = -\partial_{d} \psi_e = -\dfrac{\text{d}f_d}{\text{d}d} \dfrac{1}{2} {K} g^2 , \\
   q_d &= \partial_{\xi_d} \psi = \partial_{\xi_d} \psi_d        = h_1(1-e^{-h_2\xi_{d}}).
\end{align}
Similar to plasticity, the damage yield criterion is defined as
\begin{align}
\label{eq:dm_y}
   \phi_d(Y,q_d) &= Y - \left( Y_{0} + q_d \right).
\end{align}
In this model, the damage initiation point is controlled by the interface parameter $Y_0$, whereas $q_d$ accounts for nonlinear damage hardening. Evolution laws for the damage internal variables read:
\begin{align}
\label{eq:dmg_ev} 
   \dot{d}        &= \dot{\lambda}_d~\dfrac{\partial \phi_d}{\partial Y}    = \dot{\lambda}_d, \\
   \dot{\xi}_{d}  &= -\dot{\lambda}_d~\dfrac{\partial \phi_d}{\partial q_d} = \dot{\lambda}_d.
\end{align}
Finally, the loading/unloading conditions of the damage model are taken into account:
\begin{align}
\label{eq:dmg_kkt}
   \dot{\lambda}_d\geq0,~~~\phi_d\leq0,~~~\dot{\lambda}_d~\phi_d=0.
\end{align}

\subsection{Implementation and algorithmic aspects for the damage model}
Similar to the case for plasticity, the superindex $k$ in Algorithm~\ref{alg:dmg} indicates the $k$-th iteration in the internal loop of the damage solver. In this algorithm, $\Delta \lambda_d=\Delta t~\lambda_d$, where $\Delta t$ is the pseudo time step. When the damage is active ($\phi_d > 0$), the damage residuals (${r}^{(1)}_d$ and $r^{(2)}_d$) are linearized using the Newton-Raphson method to solve for the unknowns ($d^{k+1}$ and $\xi^{k+1}_d$).
The matrix $\boldsymbol{K}_d = \partial \boldsymbol{r}_d / \partial \boldsymbol{U}_d$ includes the derivatives of the damage residual vector $\boldsymbol{r}_d$ with respect to the unknowns ($\boldsymbol{U}_d={d^{i+1},\xi^{i+1}_d}$). If the changes in the solution of the unknown internal variables are smaller than a certain tolerance ($\text{tol}=10^{-10}$), the results are considered converged and used to compute the traction values at the interface and the tangent operator. These quantities are used to set up the global residual vector at the finite element level \cite{REZAEI2019325}.

\begin{algorithm}[H]
\caption{Solving the damage governing equations at pseudo time $t^{i+1}$}
\label{alg:dmg}
\begin{algorithmic}[1]
\item[\textbf{Input:}] interface's parameters, current gap: ${g}^{i+1}$, history variables: $d^i$, $\xi^i_d$
\item[\textbf{Output:}] current traction: ${T}^{i+1}$, current history variables $d^{i+1}$, $\xi_d^{i+1}$, tangent: ${C}_d=\text{d} {T}^{i+1} / \text{d} {g}^{i+1}$
\State $k=1$
\State Set trial values: ${g}^k_p={g}^{i}_p$, $d^k=\xi^{i}_d=d^i$
\State \textbf{loop} \textit{damage solver} \textbf{do} \\
\vspace{1mm}
$\qquad $ \textbf{if} $\phi^{tr}_d = {K} (g^{i+1})^2 - \left( Y_{0} + q_d(\xi^{i}) \right) \le 0$ \\
\vspace{1mm}
$\qquad $ $\qquad $   $d^{i+1}=d^{i}$, $\xi^{i+1}_d=\xi^{i}_d$ and ${C}_d=(1-d^{i+1})^2K $ \\ 
\vspace{1mm}
$\qquad $ \textbf{else} solve for $d^{k+1}$ and $\xi^{k+1}_d = \xi^{i}_d+\Delta \lambda^{k+1}_d$ (see Eqs~\ref{eq:dm_y},\ref{eq:dmg_ev})\\
\vspace{1mm}
$\qquad $  $\qquad  r^{(1)}_d = d^{k+1} - d^i - (\xi^{k+1}_d - \xi^{i}_d) \dfrac{\partial \phi_d}{\partial Y} \stackrel{!}{=} 0$ \\
$\qquad $  $\qquad  r^{(2)}_d = \phi_d = Y - \left( Y_{0} + q_d\left( \xi^{i+1}_d \right) \right) \stackrel{!}{=} 0$ \\
\vspace{1mm}
$\qquad$  $\qquad [d^{i+1,k+1}~~~ \xi^{i+1,k+1}_d]^T = [d^{i+1,k}~~~\xi^{i+1,k}_d]^T - \boldsymbol{K}_d^{-1} \boldsymbol{r}_d$ \\
$\qquad $ \textbf{end if} \\
\vspace{1mm}
$\qquad $ \textbf{if} $\left| {d}^{i+1,k+1}-{d}^{i+1,k}\right|>\text{tol} $ \textbf{OR} $\left| \xi^{i+1,k+1}_d-\xi^{i+1,k}_d\right|>\text{tol}$ \\
$\qquad $ $\qquad $   $k++$ \\
$\qquad $ $\qquad $   \textbf{CYCLE loop} \textit{damage solver} \\ 
$\qquad $ \textbf{else} \\
$\qquad $ $\qquad $  $d^{i+1}=d^{i+1,k+1}$, $\xi^{i+1}_d=\xi^{i+1,k+1}_d$ \\
\vspace{1mm}
$\qquad $ $\qquad $   \textbf{EXIT loop} \textit{damage solver} \\
$\qquad $ \textbf{end if}
\vspace{1mm}
\State \textbf{end loop}
\State Compute traction ${T}=(1-d^{i+1})^2 {K} {g}^{i+1}$
\State Compute tangent ${C}_d $ 
\end{algorithmic}
\end{algorithm}

\subsection{PINN to solve for local damage model or nonlinear softening behavior at the interface} \label{sec:dmg}
\begin{figure}[H] 
  \centering
  \includegraphics[width=0.99\linewidth]{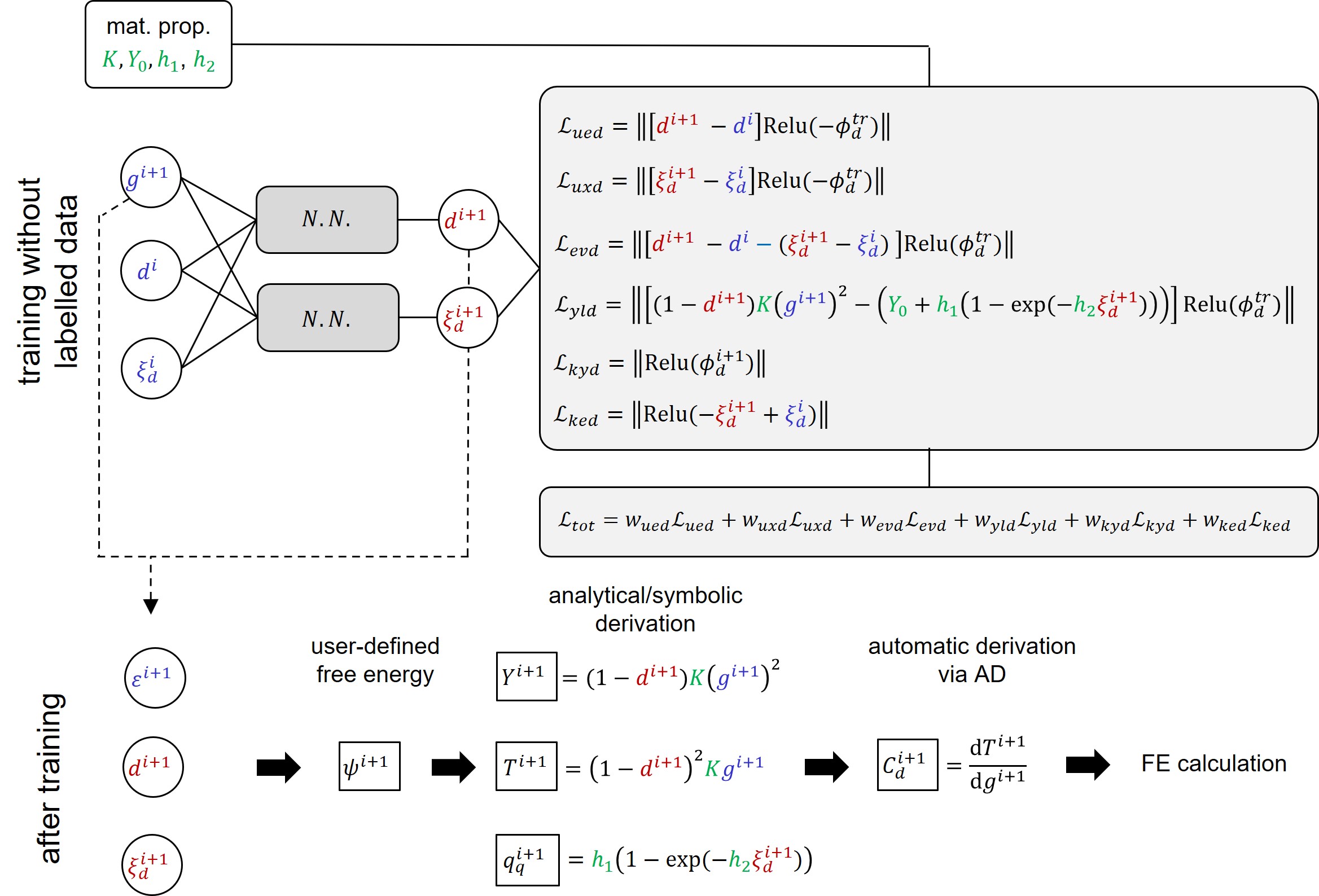}
  \caption{ Network architecture and loss functions for the COMM-PINN applied to interface failure with nonlinear softening law. }
  \label{fig:dmg_pinn}
\end{figure}

For the case of local damage evolution (interface fracture), the input layer includes the gap value $g^{i+1}$, the previous damage value $d^{i}$, and the damage hardening variable $\xi^{i}_d$. We denote the input layer via vector $\bm{X}=\{g^{i+1}, d^{i},\xi^{i}_d\}$. The output layer includes $\bm{Y}=\{d^{i+1}, \xi^{i+1}_d\}$. Similar to the previous case, we intend to use the separate fully connected feed-forward neural networks for each output variable (see also Fig.~\ref{fig:dmg_pinn}). The outcome of the constitutive material modeling via neural network reads
\begin{align}
\label{eq:NN_dmg}
    d^{i+1} = \mathcal{N}_{d} (\bm{X}; \bm{\theta}),~~~~~
    \xi^{i+1}_d = \mathcal{N}_{\xi_d} (\bm{X}; \bm{\theta}).
\end{align}

Next, we build the residuals for the introduced damage model. Here, one requires to build the damage yield function as well as evolution equations for the internal variables. These expressions will be used to construct loss functions for the neural networks. Denoting the summation of total loss terms for damage by $\mathcal{L}_{dt}$, it is defined based on the Alg.~\ref{alg:dmg}, as
\begin{align}
\label{Totalloss_dmg}
\mathcal{L}_{dt} &= 
\underbrace{w_{ued}\mathcal{L}_{ued} + w_{uxd}\mathcal{L}_{uxd}}_{\text{elastic response}} + \underbrace{w_{evd}\mathcal{L}_{evd}  + w_{yld}\mathcal{L}_{yld}}_{\text{damage evolution}} + \underbrace{w_{ked}\mathcal{L}_{ked}  + w_{kyd}\mathcal{L}_{kyd}}_{\text{KKT conditions}}
\end{align}
In Eq.\,(\ref{Totalloss_dmg}), different loss terms cover all the possible loading, unloading, and reloading scenarios. The first two terms ($\mathcal{L}_{ued}$ and $\mathcal{L}_{uxd}$) guarantee that there is no evolution of damage when the trial yield function ($\phi^{tr}_d$) is negative. Once $\phi^{tr}_d>0$, the main damage residuals for evolution law, as well as the current yield function, become active which is denoted by $\mathcal{L}_{evd}$ and $\mathcal{L}_{yld}$, respectively. Finally, to make sure that the KKT conditions are always satisfied, we have the last two loss terms ($\mathcal{L}_{ked}$ and $\mathcal{L}_{kyd}$). All these relevant loss terms are summarized in what follows
\begin{align}
\label{loss_ued}
\mathcal{L}_{ued} &= \text{MSE}\left( (d^{i+1}-d^{i})\text{Relu}(-\phi^{tr}_d) \right), \\
\label{loss_uxd}
\mathcal{L}_{uxd} &= \text{MSE}\left( (\xi^{i+1}_d-\xi^{i}_d)\text{Relu}(-\phi^{tr}_d) \right), \\
\label{loss_evd}
\mathcal{L}_{evd} &= \text{MSE}\left( (d^{i+1}-d^{i}-(\xi^{i+1}_d-\xi^{i}_d))\text{Relu}(\phi^{tr}_d) \right), \\
\label{loss_yld}
\mathcal{L}_{yld} &= \text{MSE}\left( \left( (1-d^{i+1}) K (g^{i+1})^2 - ( Y_{0} + h_1(1-\text{exp}(-h_2\xi^{i+1}_d)) \right)\text{Relu}(\phi^{tr}_d) \right), \\
\label{loss_ked}  
\mathcal{L}_{ked} &= \text{MSE}\left( \text{Relu}(\phi^{i+1}_d) \right), \\
\label{loss_kyd}
\mathcal{L}_{kyd} &= \text{MSE} \left( \text{Relu}(-\xi^{i+1}_d+\xi^{i}_d) \right).
\end{align}
The so-called trial yield function $\phi^{tr}_d$ is evaluated by means of the quantities at the input layer (see also Alg.~\ref{alg:dmg}):
\begin{align}
\label{trial_dmg}
\phi^{tr}_d = (1-d^{i}) K g^{i+1} - \left( Y_0 + h_1(1-\text{exp}(-h_2\xi^{i}_d)) \right).
\end{align}

In Fig.~\ref{fig:dmg_pinn}, we summarized the main loss terms for the interface fracture model with nonlinear softening behavior. For the sake of clarity, all the input variables are denoted in blue color while the output (unknown) variables are presented in red color. Note that the material (interface) properties are kept constant and represented by green color.

After the training is completed, the network predicts the unknown internal variables (i.e. $d^{i+1}$ and $\xi^{i+1}_d$). Having the updated internal variables as well as the current gap value ($g^{i+1}$), one can construct the free energy function as well as the traction value ($T^{i+1}$). Note that the functionality of traction is obtained by analytical differentiation of the energy function with respect to the gap. Finally, the tangent operator which is defined as the full derivative of the current traction value with respect to the current gap value is obtained via automatic differentiation.

The final loss term is minimized at every single collocation point. The mathematical optimization problem is written as  
\begin{align}
\label{minimize_2}
\bm{\theta}^* = \arg \min_{\bm{\theta}} \mathcal{L}_{dt}(\bm{X}; \bm{\theta}),
\end{align}
where $\bm{\theta}^*$ are the optimal trainable parameters (weights and biases) of the network.

\section{Generation of collocation points}
The location of collocation points plays a crucial role in training \cite{Henkes2022, REZAEI2022PINN}. \textit{Collocation points} are initial inputs for which the governing equations are going to be satisfied (collected). The location of collocation points can be different than \textit{data points} for which we know the solution in advance. 
The inputs of the NN are the given strain or gap at step $i+1$ (i.e. $\varepsilon^{i+1}$ or $g^{i+1}$) as well as state variables from the previous time step $i$ (i.e. $\varepsilon^{i}_p$ and $\xi_p^{i}$ or $d^{i}$ and $\xi_d^{i}$). 

In this approach, we propose to assign admissible values to the set of inputs. However, two aspects should be considered when generating these points: 1) the desired range of the inputs and 2) avoiding any violation of important physical aspects when assigning random values. For instance, in the case of elastoplasticity, the accumulative plastic strain is always positive and greater than the plastic strain. Similarly, in the case of interface damage and the local damage model, the damage variable and damage hardening variable are always positive, below one, and identical. These restrictions are typically well-known and straightforward to account for. However, for more complex material models, simpler strategies will be required for future developments. It is essential to ensure that the generated values are physically consistent and within the desired range to prevent issues during training and improve the accuracy of the model.


Another idea is to generate random loading paths and gather the data from material models as initial values and substitute these models with the trained NN. For the current work we have tried both of these options and concluded that with both of these methods, one is able to solve the constitutive material equations. For the sake of simplicity, we report the more straightforward approach for the generation of collocation points as explained in the following algorithms.

For the case of elastoplastic behavior, readers are referred to Alg.~\ref{alg:data_gen_plas}. Here, we are using the following intervals for collocation points generation: $bg_e=0$, $st_e=0.01$, $en_e=+1$, $bg_{\varepsilon_p}=0$, $st_{\varepsilon_p}=0.01$, $en_{\varepsilon_p}=+1$, $bg_{\xi_p}=0$, $st_{\xi_p}=0.01$, and $en_{\xi_p}=+1$. The main outcome of the Alg.~\ref{alg:data_gen_plas} is the location of collocation points stored in the matrix $\bold{IN}_p$ which has three columns for the three main inputs of the NN. The number of rows depends on the user's choice of intervals and steps. 
%
\begin{algorithm}[H]
\caption{Collocation point generation for solving elasto-plastic constitutive relations}
\label{alg:data_gen_plas}
\begin{algorithmic}[1]
\item[\textbf{Input:}] step size: $st$, begin $bg$ and end of interval $en$
\item[\textbf{Output:}] admissible set of inputs for training $\bold{IN}_p$
\State $n=1$
\State Initializing: $\boldsymbol{\varepsilon}=[bg_{\varepsilon}:st_{\varepsilon}:en_{\varepsilon}]$, $\boldsymbol{\varepsilon}_p=[bg_{\varepsilon_p}:st_{\varepsilon_p}:en_{\varepsilon_p}]$, $\boldsymbol{\xi}_p=[bg_{\xi_p}:st_{\xi_p}:en_{\xi_p}]$
\State \textbf{for} l \textbf{in} $\boldsymbol{\varepsilon}$ \textbf{do} \\
\vspace{1mm}
$\qquad $ \textbf{for} j \textbf{in} $\boldsymbol{\varepsilon}_p$ \textbf{do} \\
\vspace{1mm}
$\qquad $ $\qquad $ \textbf{for} k \textbf{in} $\boldsymbol{\xi}_p$ \textbf{do} \\ 
\vspace{1mm}
$\qquad $ $\qquad $ $\qquad $ $\bold{IN}_p[n,1]$ = $\boldsymbol{\varepsilon}[l]$\\
\vspace{1mm}
$\qquad $ $\qquad $ $\qquad $ $\bold{IN}_p[n,2]$ = $\boldsymbol{\varepsilon}_p[j]$\\
\vspace{1mm}
$\qquad$ $\qquad$ $\qquad$ \textbf{if} $\boldsymbol{\xi}_p[k] \ge \boldsymbol{\varepsilon}_p[j]$  \\
\vspace{1mm}
$\qquad $ $\qquad$ $\qquad$ $\qquad$ $\bold{IN}_p[n,3]$ = $\boldsymbol{\xi}_p[k]$\\
\vspace{1mm}
$\qquad $ $\qquad $ $\qquad$ $\qquad$ $n++$\\
$\qquad $ $\qquad $ $\qquad$ \textbf{end if}
\end{algorithmic}
\end{algorithm}

For the case of the interface damage model, readers are referred to Alg.~\ref{alg:data_gen_dmg}. We are using following intervals for collocation points generation: $bg_g=0$, $st_g=0.02$, $en_g=+1$, $bg_{d}=0$, $st_{d}=0.02$, $en_{d}=+1$, $bg_{\xi_d}=0$, $st_{\xi_p}=0.02$, and $en_{\xi_d}=+1$. The main outcome of the Alg.~\ref{alg:data_gen_dmg} is the location of collocation points for training the damage model stored in the matrix $\bold{IN}_d$. 
\begin{algorithm}[H]
\caption{Collocation point generation for solving the interface damage model}
\label{alg:data_gen_dmg}
\begin{algorithmic}[1]
\item[\textbf{Input:}] step size: $st$, begin $bg$ and end of interval $en$
\item[\textbf{Output:}] admissible set of inputs for training $\bold{IN}_d$
\State $n=1$
\State Initializing: $\boldsymbol{g}=[bg_{g}:st_{g}:en_{g}]$, $\boldsymbol{d}=[bg_{d}:st_{d}:en_{d}]$, $\boldsymbol{\xi}_d=[bg_{\xi_d}:st_{\xi_d}:en_{\xi_d}]$
\State \textbf{for} l \textbf{in} $\boldsymbol{g}$ \textbf{do} \\
\vspace{1mm}
$\qquad $ \textbf{for} j \textbf{in} $\boldsymbol{d}$ \textbf{do} \\
\vspace{1mm}
$\qquad $ $\qquad $ \textbf{for} k \textbf{in} $\boldsymbol{\xi_d}$ \textbf{do} \\ 
\vspace{1mm}
$\qquad $ $\qquad $ $\qquad $ $\bold{IN}_d[n,1]$ = $\boldsymbol{g}[l]$\\
\vspace{1mm}
$\qquad $ $\qquad $ $\qquad $ $\bold{IN}_d[n,2]$ = $\boldsymbol{d}[j]$\\
\vspace{1mm}
$\qquad$ $\qquad$ $\qquad$ \textbf{if} $\boldsymbol{\xi}_d[k] = \boldsymbol{d}[j]$  \\
\vspace{1mm}
$\qquad $ $\qquad$ $\qquad$ $\qquad$ $\bold{IN}_d[n,3]$ = $\boldsymbol{\xi}_d[k]$\\
\vspace{1mm}
$\qquad $ $\qquad $ $\qquad$ $\qquad$ $n++$\\
$\qquad $ $\qquad $ $\qquad$ \textbf{end if}
\end{algorithmic}
\end{algorithm}

\section{Results}
The algorithms developed in the current work are implemented in the SciANN package \cite{SciANN} and the methodology can be easily transferred to other programming platforms. For all the reported results, the Adam optimizer is employed. We start with the results of the damage model due to its simplicity compared to the case of plasticity. Since the two models share some similarities, the majority of the hyperparameter studies are only reported for one of these models to avoid unnecessary repetition.

\subsection{Case studies on the interface damage model} 
The relevant network parameters for learning the damage model at the interface are summarized in Table\,\ref{tab:network_dmg}. Some of these parameters are obtained after extensive (hyper) parameter studies. To make their influence clear to the readers, we will change these values and functions to investigate their impact on the obtained results.
\begin{table}[H]
\centering
\caption{Summary of the COMM-PINN network parameters for the damage model.}  
\label{tab:network_dmg}
\begin{footnotesize}
\begin{tabular}{ l l }
\hline
Parameter                          &  Value    \\
\hline
Input, Output                  &   $\{g^{i+1}, d^{i},\xi^{i}_d\}$, $\{d^{i+1},\xi^{i+1}_d\}$ \\ 
Activation function                & Relu, Tanh, Sigmoid  \\ 
Number of layers and neurons per layer ($L$, $N_l$)  &  (5, 10),~~~(5, 50),~~~(5, 100)  \\
Batch size                         &  500  \\ 
Learning rate $\alpha$, number of epochs  &  $(10^{-4},10^{3})$  \\ 
\hline
\end{tabular}
\end{footnotesize}
\end{table} 

The material parameters reported in Table \ref{tab:dmg_parameters} are chosen in a way to keep the stress values below value one and they do not necessarily represent any realistic material at this point. Nevertheless, one can calibrate the model based on some other measurements (see \cite{REZAEI2019325}) and normalize the model variables to keep them in the range between $-1$ and $1$. As we checked, the performance of the proposed methodology did not change for other chosen material properties. 
\begin{table}[H]
\caption{Material parameters for the interface damage model described in section \ref{sec:dmg}.}
	\label{tab:dmg_parameters}
	\centering
	\begin{tabular}{ l l l } \hline
	\multirow{1}{*}{}                      & Unit                      & Value                \\ \hline \hline
	Normal interface stiffness $K$       & [MPa/mm]              & $5.0$      \\ 
	Damage initiation criterion $Y_{0}$  & [MPa mm]                     & $0.1$     \\
	Damage hardening parameter $h_1$     & [MPa mm]                       & $2.0$                \\
	Damage hardening parameter $h_2$     & [$1/$mm]                & $1.0\times10^{2}$
    \\ \hline
	\end{tabular}
\end{table}

\begin{figure}[H] 
  \centering
  \includegraphics[width=0.8\linewidth]{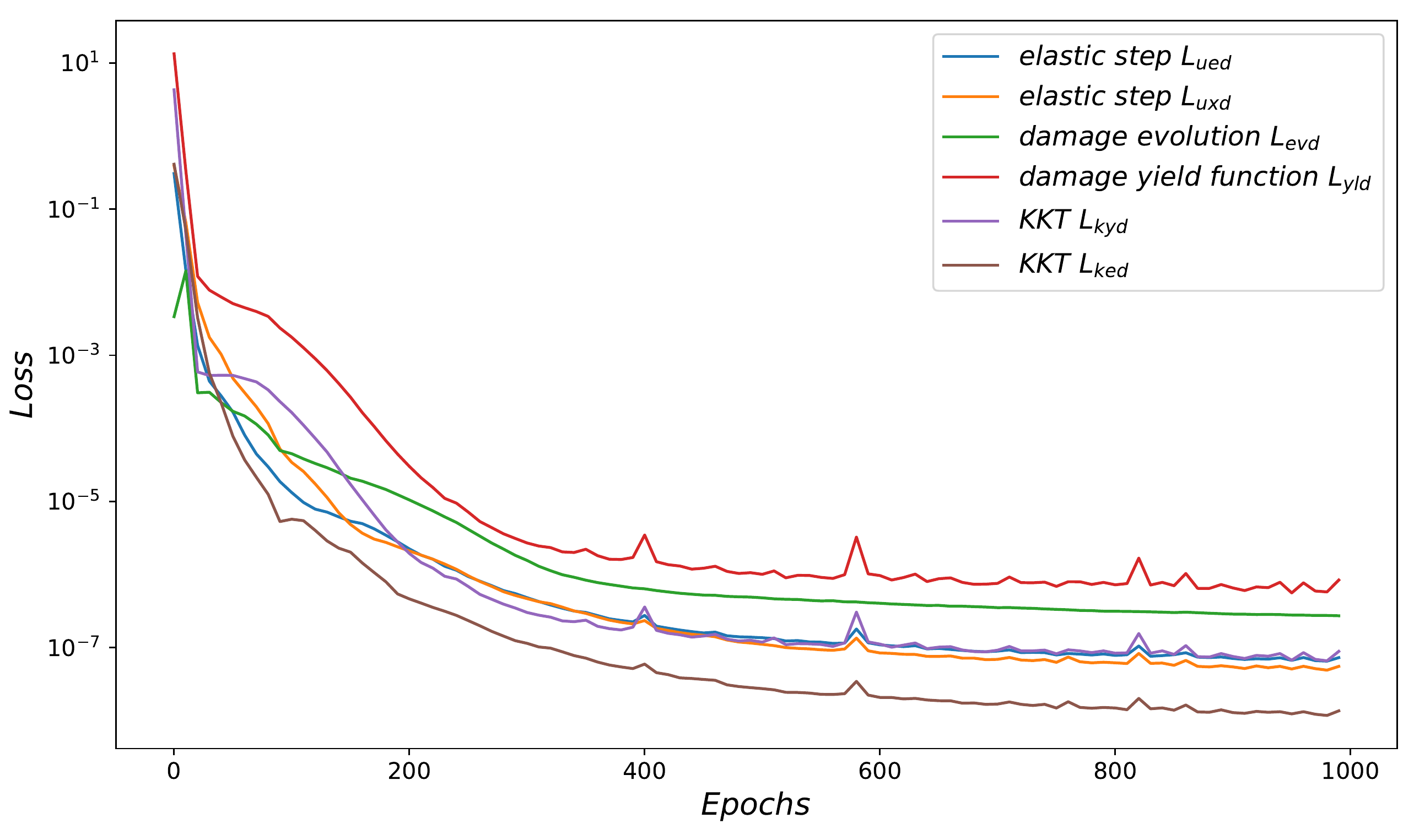}
  \caption{Prominent loss terms for the interface damage model with nonlinear softening law utilizing the proposed COMM-PINN algorithm.}
  \label{fig:loss_dmg}
\end{figure}

\begin{figure}[H] 
  \centering
  \includegraphics[width=0.99\linewidth]{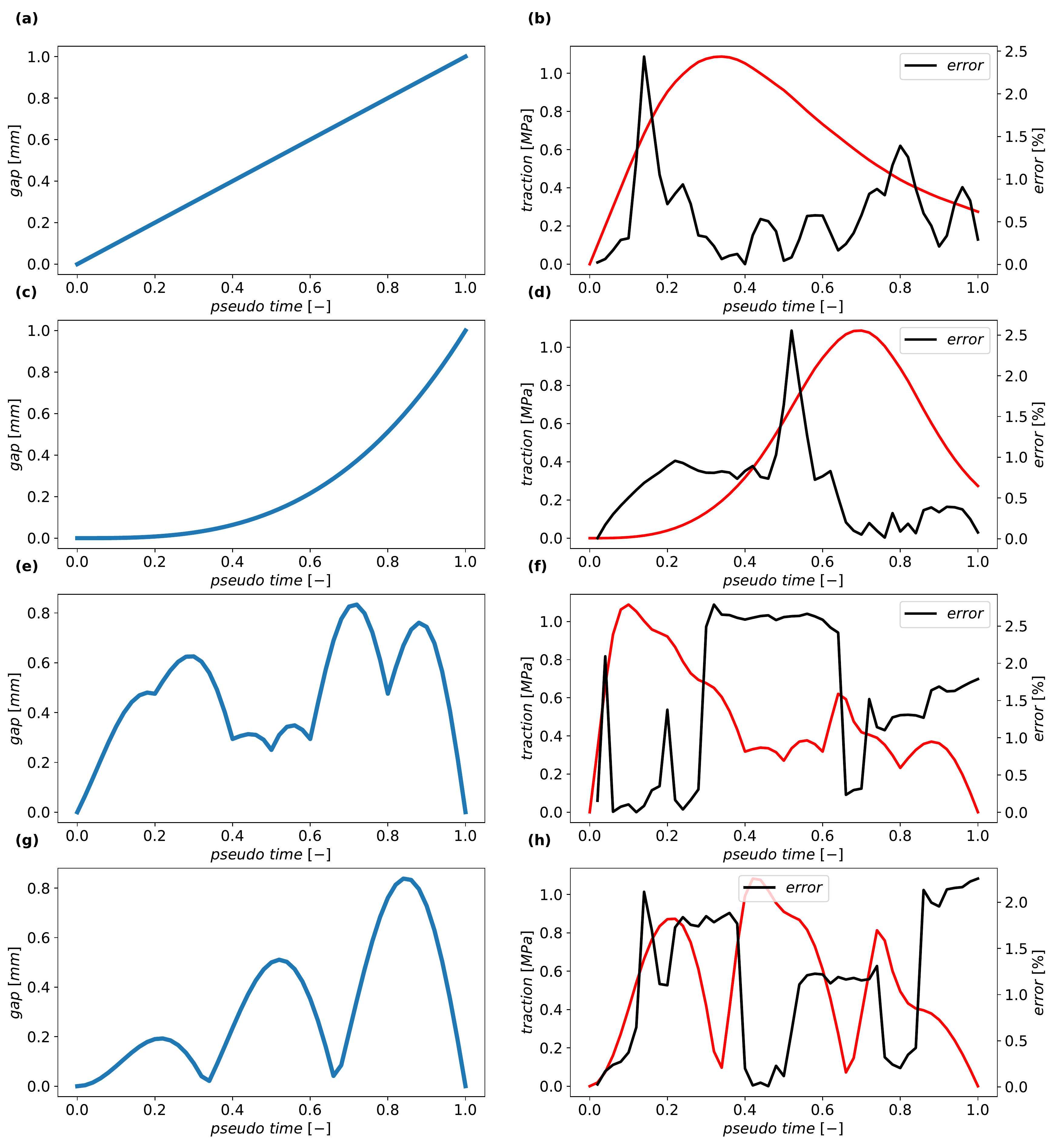}
  \caption{Predictions of the trained COMM-PINN for the interface damage model with nonlinear softening law.}
  \label{fig:damage_error}
\end{figure}

The evolution of each loss term is shown in Fig.~\ref{fig:loss_dmg} as a function of epochs. The collocation points for the training are based on the last section. We used equal weightings for all the loss terms at the beginning and it turned out that the prediction of the NN model is acceptable by doing so (i.e. we have $w_{ued}=w_{uxd}=w_{evd}=w_{yld}=w_{ked}=w_{kyd}=1.0$). Furthermore, for the results provided in Fig.~\ref{fig:loss_dmg}, we utilize Relu as the activation function and $5$ hidden layers with $100$ neurons in each. The rest of the parameters are according to Table~\ref{tab:network_dmg}. All the loss functions decay simultaneously and we did not notice any significant improvement after $1000$ epochs. We also observed the same response for all the loss terms through different hyper-parameter studies. 
In what follows, we will report the performance of the trained NN versus the response we obtained based on the classical material model routine (i.e. return mapping algorithm) which is our reference solution. Since the NN has not trained based on any given data (i.e. known and available solution), any arbitrary loading path we define for the strain value can be seen as a test case to evaluate its performance. Nevertheless, as we go through the results section, we examine the performance of the NN beyond the range of the collocation points. Some early predictions for various loading paths are shown in Fig.~\ref{fig:damage_error}. Here, the first column represents the input gap value in time (i.e. $g(t)$), and the second column represents the obtained traction in time (i.e. $T(t)$). For the first and second rows, we applied $g(t) = t$ and $g(t) = t^3$, respectively. For the third and fourth rows, we have $g(t) = 0.5 |t~\text{sin}(5\pi t)| + 0.5 |\text{sin}(2\pi t)|$ and $g(t) = 1.0 |t~\text{sin}(3\pi t)|$, respectively. 

To evaluate the NN performance, we also over-plot the relative error for the predicted traction values. The error $err$ is calculated via a point-wise comparison of the traction from NN $T_N$ and calculated traction from the standard return mapping algorithm $T_M$. Therefore, we have $err = \frac{|T_N-T_M|}{T_M}\times 100$. The average error for different loading scenario is about $1\%$ and in some extreme cases the maximum error goes up to $2.5\%$. We will discuss possible suggestions to improve the NN predictions through this section.

To avoid repetition, in rest of the paper, we chose a rather complicated loading path for all the studies which includes several loading, unloading, and reloading scenarios. Consequently, we can make sure that all the possible scenarios can become active. For the following studies, we force the gap vector to change through time via the following equation $g(t) = 2.0 |t~\text{sin}(3\pi t)|$. In this loading case, the gap will go beyond the value of $1.5$mm while we only train the NN with collocation points up to a gap value of $1.0$. Therefore, toward the end of the loading path, we have been checking the extrapolation capabilities of the trained NN.

Note that even in the case of PINN without any initial data, the location of collocation points has a big impact on the final performance. In other words, by going beyond the range of the collocation points, there is no guarantee about the acceptable predictions of the NN. 

\subsubsection{Influence of neuron's number in NN architecture}
\begin{figure}[H] 
  \centering
  \includegraphics[width=0.99\linewidth]{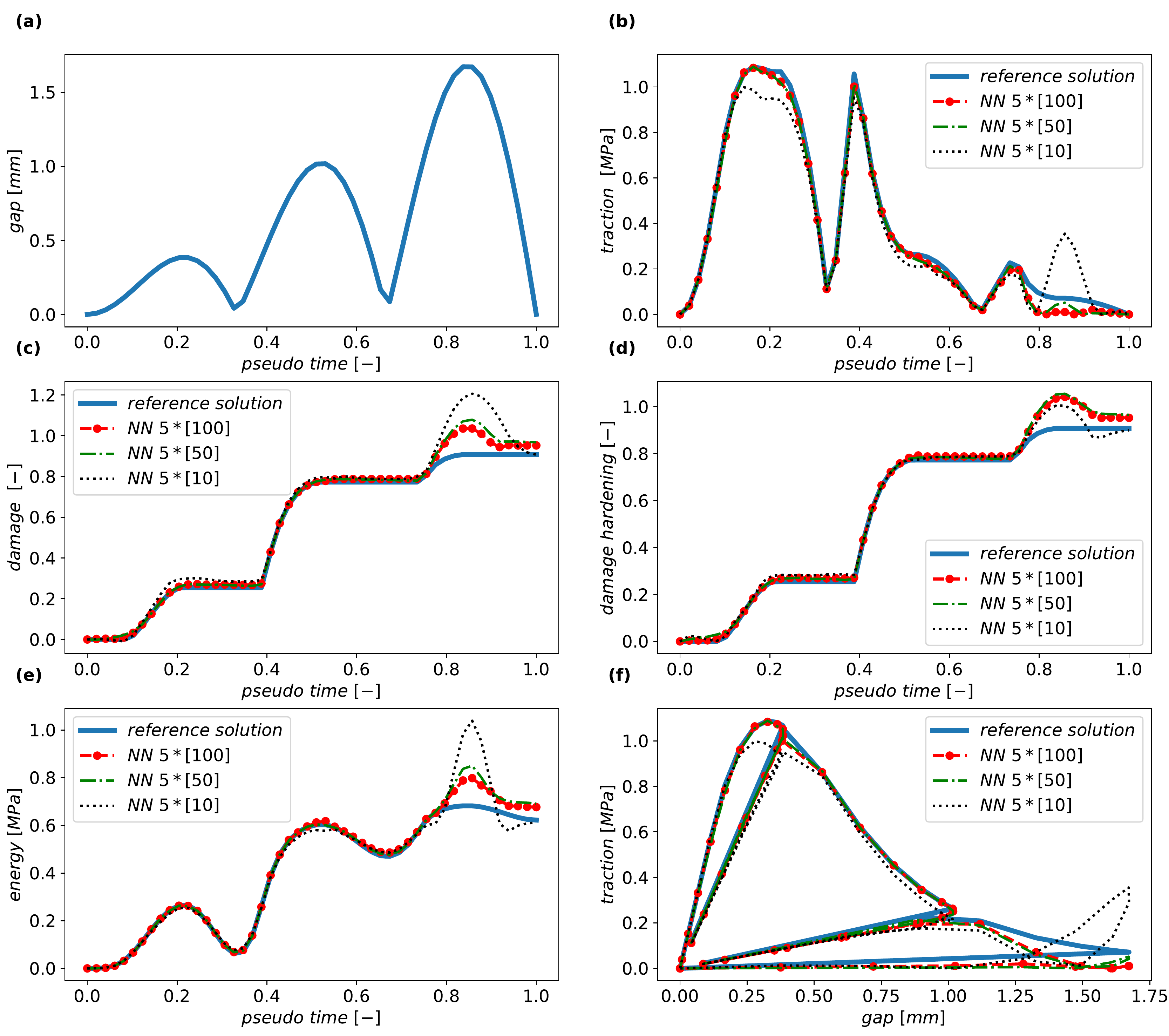}
  \caption{Influence of the NN layers on the prediction of the network. (a) the loading (gap over time) is based on $g(t) = 2.0 |t~\text{sin}(3\pi t)|$. (b) Predicted traction $T(t)$ via train NN over time. (c) predicted damage variable $d(t)$ via train NN over time. (d) predicted damage hardening variable $\xi_d(t)$ via train NN over time. (e) predicted free energy function $\psi_d(t)$ over time. (f) predicted traction versus the gap value. }
  \label{fig:non_damage}
\end{figure}

In Fig.~\ref{fig:non_damage}, we study the influence of various NN architecture structures. Here collocation points are generated with step size $0.02$ and the Relu activation function is utilized. By comparing different NNs, we conclude that one requires enough neurons to capture all the complicated nonlinear behavior. At some point, adding extra neurons or layers is not beneficial anymore and might even result in the overfitting of the model. Similar performance was also observed for the number of layers where upon choosing a very small number of layers, the performance is really poor. Interestingly enough, for the red curve where we used a proper architecture, the NN can accurately capture the nonlinear response for all the loading, unloading, and reloading test cases. Towards the end of the loading, where we go beyond the range of the provided collocation points for the training, we observe some deviations in the predicted response. In other words, even thermodynamical constraints can be violated when it comes to evaluating the NN response beyond the range of collocation points. On the positive side, one can always increase the range and density of the collocation points easily, unlike pure data-driven methods for which we need to obtain new data for the new range. It should be noted that using more collocation points will increase the training time. A fair comparison between pure physics-driven and pure data-driven NN training is not so trivial as they are different in the nature of loss functions and initial data sets and computational costs. Nevertheless, in this work we will provide some initial comparisons to observe the deviation for a purely data-driven approach when it comes to extrapolation.

\subsubsection{Influence of activation function}
\begin{figure}[H] 
  \centering
  \includegraphics[width=0.99\linewidth]{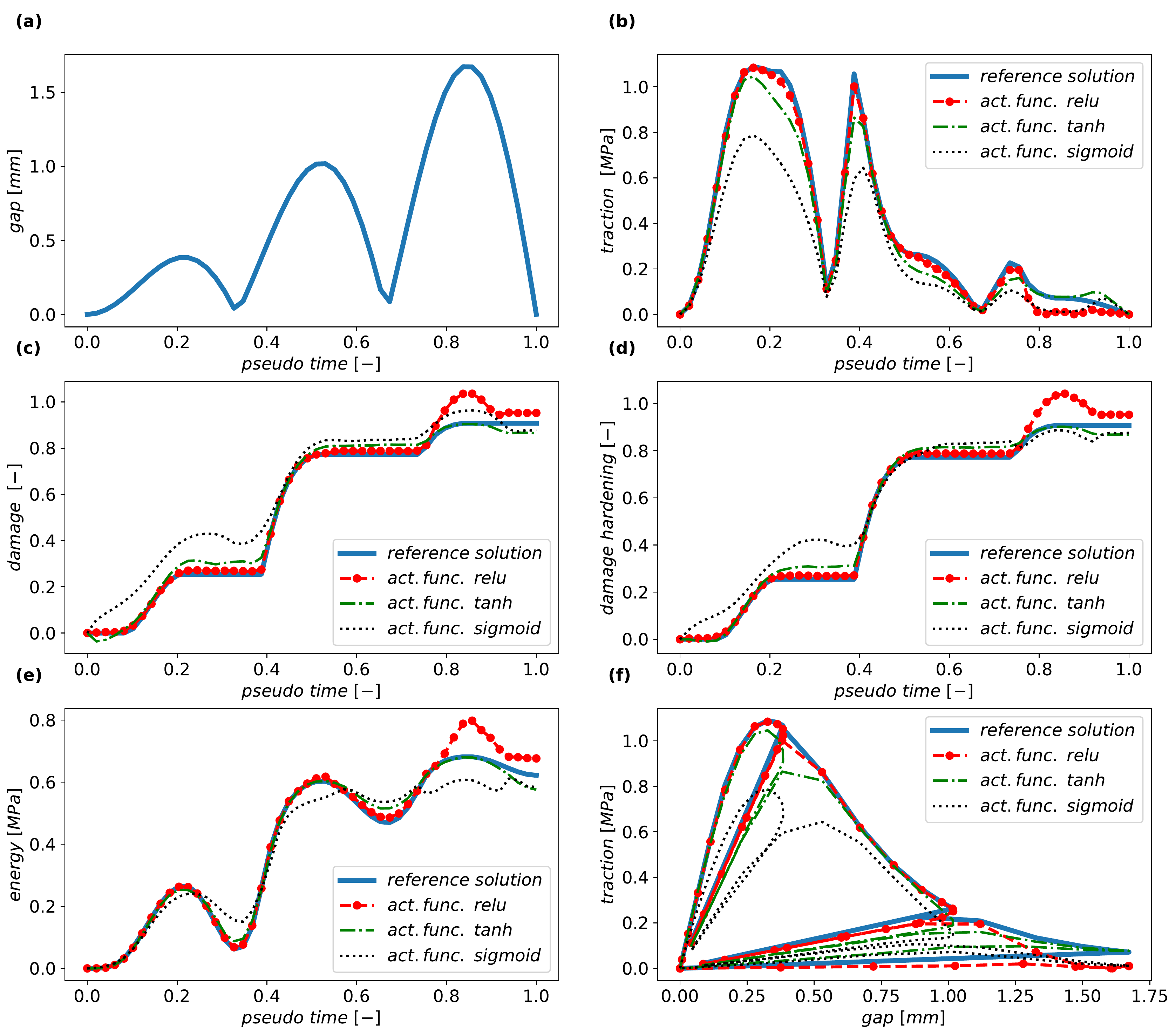}
  \caption{Influence of the choice of activation function on the prediction of the network. (a) the loading (gap over time) is based on $g(t) = 2.0 |t~\text{sin}(3\pi t)|$. (b) Predicted traction $T(t)$ via train NN over time. (c) predicted damage variable $d(t)$ via train NN over time. (d) predicted damage hardening variable $\xi_d(t)$ via train NN over time. (e) predicted free energy function $\psi_d(t)$ over time. (f) predicted traction-separation relation.}
  \label{fig:activation_damage}
\end{figure}

In Fig.~\ref{fig:activation_damage}, we look into the influence of various choices for the activation functions on the predictions. The explanations of the initial training set are similar to those before. Here 5 layers with 100 neurons in each are utilized for all the case studies. We concluded that the Relu function shows the best performance compared to others. Although not reported, our conclusion is the same even by using other activation functions such as Swish and Softplus. The better performance for the Relu function is perhaps due to its nature with a sharp transition that can capture the sudden change of behavior in the unloading, reloading, and yield condition.

\subsubsection{Influence of collocation point density}
\begin{figure}[H] 
  \centering
  \includegraphics[width=0.99\linewidth]{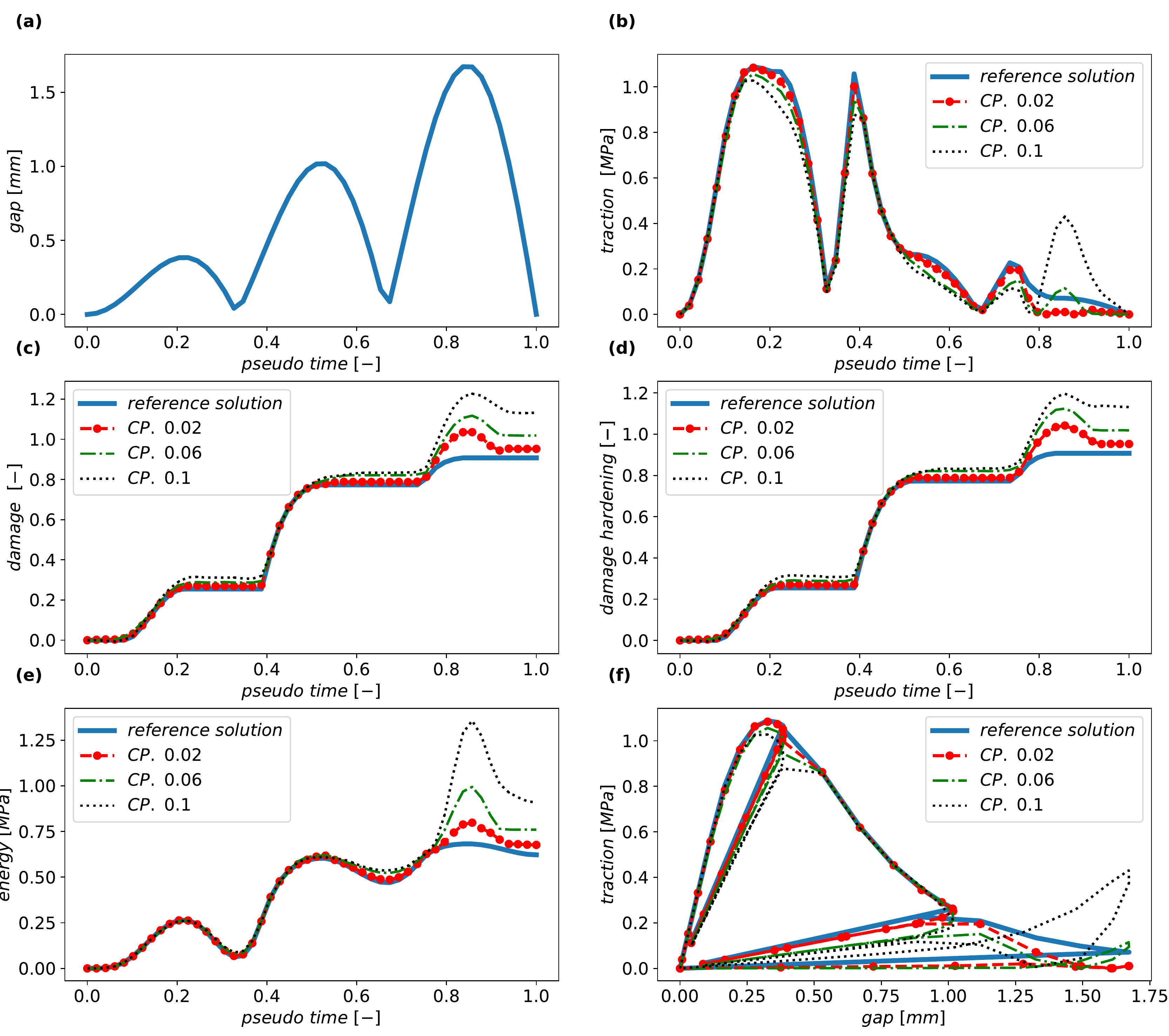}
  \caption{Influence of the step size for generation of collocation points on the predictions. (a) the loading (gap over time) is based on $g(t) = 2.0 |t~\text{sin}(3\pi t)|$. (b) Predicted traction $T(t)$ via train NN over time. (c) predicted damage variable $d(t)$ via train NN over time. (d) predicted damage hardening variable $\xi_d(t)$ via train NN over time. (e) predicted free energy function $\psi_d(t)$ over time. (f) predicted traction versus the gap value.}
  \label{fig:step_damage}
\end{figure}

In Fig.~\ref{fig:step_damage}, we look into the influence of various choices for the density of the collocation points on the predictions. The chosen activation function for this study is the Relu function and the network has 5 layers with 100 neurons. The density of the collocation points is varied from $0.1$ to $0.02$. We concluded that even with a very low number of points, the NN can capture the overall behavior of such a nonlinear loading path. We also observe that the performance of the NN model is improved by feeding more collocation points to the NN for the training process. Furthermore, having more points will improve the extrapolation capabilities of the NN (see the obtained response after time 0.8). On the other hand, more collocation points will significantly increase the computational cost of the training. Training with step size 0.02 takes twice the time compared to step size 0.1. One may have to adapt the architecture of the NN adequately based on the number of collocation points.

It should be noted that in all the cases, the test loading path is generated with 50 points in time (i.e. the time step size is $0.02$). As we will discuss this further in the next section, one can choose different step sizes and call the same NN and yet have acceptable performance.

\color{black}
\newpage
\subsection{Extension to a 3D case: anisotropic potential-based cohesive zone model}
\begin{figure}[H] 
  \centering
  \includegraphics[width=0.8\linewidth]{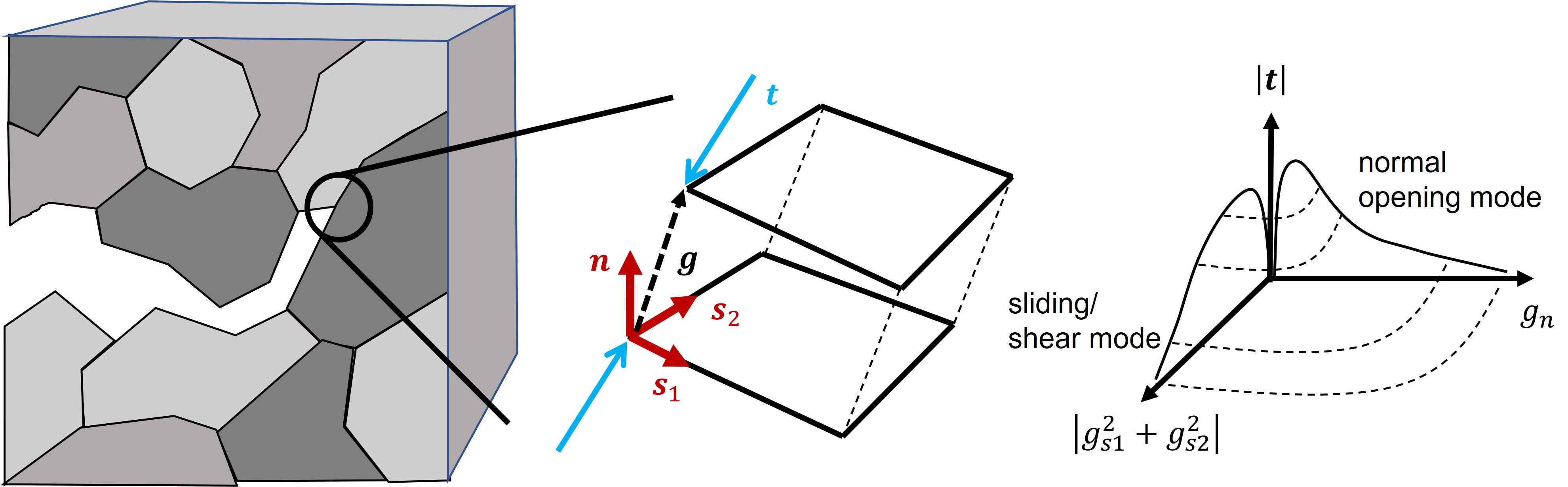}
  \caption{Intergranular fracture in polycrystalline material modeled by means of a 3D cohesive zone element.}
  \label{fig:interface_cze}
\end{figure}

Based on \cite{REZAEI2019325} and the references therein, we provide a summary of the 3D model as follows.
Figure~\ref{fig:interface_cze} illustrates a well-known intergranular damage mode observed at the microscale in polycrystalline materials. To capture this fracture behavior, the interface between all the grains, commonly referred to as grain boundaries, is enriched with classical cohesive zone elements. The displacement jump or gap vector $\boldsymbol{g}$ between two arbitrary surfaces is calculated based on the positions of CZ element nodes on these surfaces.
The normal direction $n$ and two shear directions $s_1$ and $s_2$ are defined based on the position of the mid-plane. In a 3D configuration, the gap vector $\boldsymbol{g}$ and traction vector $\boldsymbol{t}$ each have three components. Hence, we have: 
\begin{align}
\label{eq:gap_add}
   \boldsymbol{g}^T &= \left[ g_{s1},~g_{s1}~ g_n \right].\\
   \boldsymbol{t}^T &= \left[ t_{s1},~t_{s1}~ t_n \right].
\end{align}
Damage at the interface is assumed to be isotropic, and as a result, a scalar parameter $d$ is introduced as an internal variable to represent the loss of bonding force at a specific interface. The Helmholtz free energy $\psi$ at the interface comprises two components: the elastic part ($\psi_e$) and the damage hardening part ($\psi_d$). This can be expressed as follows:
\begin{equation}
\label{eq:tot_en}
   \psi = \psi_e \left( \boldsymbol{g},d \right) 
          + \psi_d \left( \xi_{d} \right) = (1-d)^2\dfrac{1}{2} (\boldsymbol{g})^T \boldsymbol{K} \boldsymbol{g}
          + h_1 \left( \xi_{d} + \dfrac{e^{-h_2\xi_{d}}-1}{h_2}  \right).
\end{equation}
In the expression above, $\xi_{d}$ represents the damage (isotropic) hardening variable, and $h_1$ and $h_2$ are the interface damage hardening parameters. The parameter $\boldsymbol{K} = K_{s_1}\boldsymbol{s_1}\otimes \boldsymbol{s_1} + K_{s_2}\boldsymbol{s_2}\otimes \boldsymbol{s_2} + K_n~\boldsymbol{n}\otimes \boldsymbol{n}$ is a second-order tensor that represents the anisotropic initial stiffness of the interface.
The thermodynamic conjugate forces associated with the interface are as follows:
\begin{align}
\label{eq:tra}
   \boldsymbol{t} &= \partial_{\boldsymbol{g}} \psi   = (1-d)^2 \boldsymbol{K} \boldsymbol{g} ,\\
\label{eq:Y}
   Y              &= -\partial_{d} \psi                 = (1-d) (\boldsymbol{g})^T \boldsymbol{K} \boldsymbol{g}  = (1-d) \left( K_{s_1} g^2_{s_1} + K_{s_2} g^2_{s_2} + K_{n} g^2_{n}\right). 
\end{align}
The relations for the yield function (Eq.~\ref{eq:dm_y}) and the evolution equation (Eq.~\ref{eq:dmg_ev}) remain unchanged. Additionally, the expressions for the loss terms are given by Eqs.\ref{loss_ued} to \ref{loss_kyd}. It is important to note that for other models, such as a 3D plasticity model, additional loss terms need to be defined for each component of plastic strain. However, this task is relatively straightforward, as one simply expresses the equations in terms of the loss term, which closely resembles the approach taken in the 1D example.

The network parameters used for training the 3D cohesive zone model are summarized in Table,\ref{tab:network_cz}. These parameters were determined through previous (hyper)parameter studies. It is worth noting that many of the neural network parameters were transferable from the simplified 1D analysis, allowing for a consistent and efficient learning process.
Please refer to Table,\ref{tab:network_cz} for a detailed overview of the network parameters employed in this study.
\begin{table}[H]
\centering
\caption{Summary of the COMM-PINN network parameters for the 3D cohesive zone model.}  
\label{tab:network_cz}
\begin{footnotesize}
\begin{tabular}{ l l }
\hline
Parameter                          &  Value    \\
\hline
Input, Output                  &   $\{g_{s1}^{i+1},g_{s1}^{i+1},g_n^{i+1}, d^{i},\xi^{i}_d\}$, $\{d^{i+1},\xi^{i+1}_d\}$ \\ 
Activation function                & Relu  \\ 
Number of layers and neurons per layer ($L$, $N_l$)  &  (5, 100)  \\
Batch size  &  50  \\ 
Learning rate $\alpha$, number of epochs  &  $(10^{-4},10^{3})$  \\ 
\hline
\end{tabular}
\end{footnotesize}
\end{table} 

The material parameters listed in Table \ref{tab:cz_parameters} were carefully selected to capture the anisotropic behavior associated with various fracture modes. During our investigations, we found that the performance of the proposed methodology remained consistent across different choices of material properties which indicates the robustness of the approach.
Refer to Table \ref{tab:cz_parameters} for a comprehensive overview of the chosen material parameters and their significance in capturing the anisotropic behavior of the system. 
\begin{table}[H]
\caption{Material parameters for the interface damage model described in section \ref{sec:dmg}.}
	\label{tab:cz_parameters}
	\centering
	\begin{tabular}{ l l l } \hline
	\multirow{1}{*}{}                      & Unit                      & Value                \\ \hline \hline
	Normal stiffness (mode I opening) $K_n$       & [MPa/mm]              & $5.0$      \\
    Shear stiffness $K_{s1}$ (mode II sliding) & [MPa/mm]              & $0.5$      \\
    Shear stiffness $K_{s2}$  (mode III sliding)     & [MPa/mm]              & $2.0$      \\
	Damage initiation criterion $Y_{0}$  & [MPa mm]                     & $0.1$     \\
	Damage hardening parameter $h_1$     & [MPa mm]                       & $2.0$                \\
	Damage hardening parameter $h_2$     & [$1/$mm]                & $1.0\times10^{2}$
    \\ \hline
	\end{tabular}
\end{table}

The collocation points used for training are based on the methodology described in the previous section. The only modification is the inclusion of additional collocation points for the two shear modes to capture the full 3D behavior. It is worth noting that the loss functions associated with the yield criteria need to be updated accordingly. However, the damage evolution equations remain the same as in the previous section. See also Table~\ref{tab:network_cz}.

\begin{figure}[H] 
  \centering
  \includegraphics[width=0.99\linewidth]{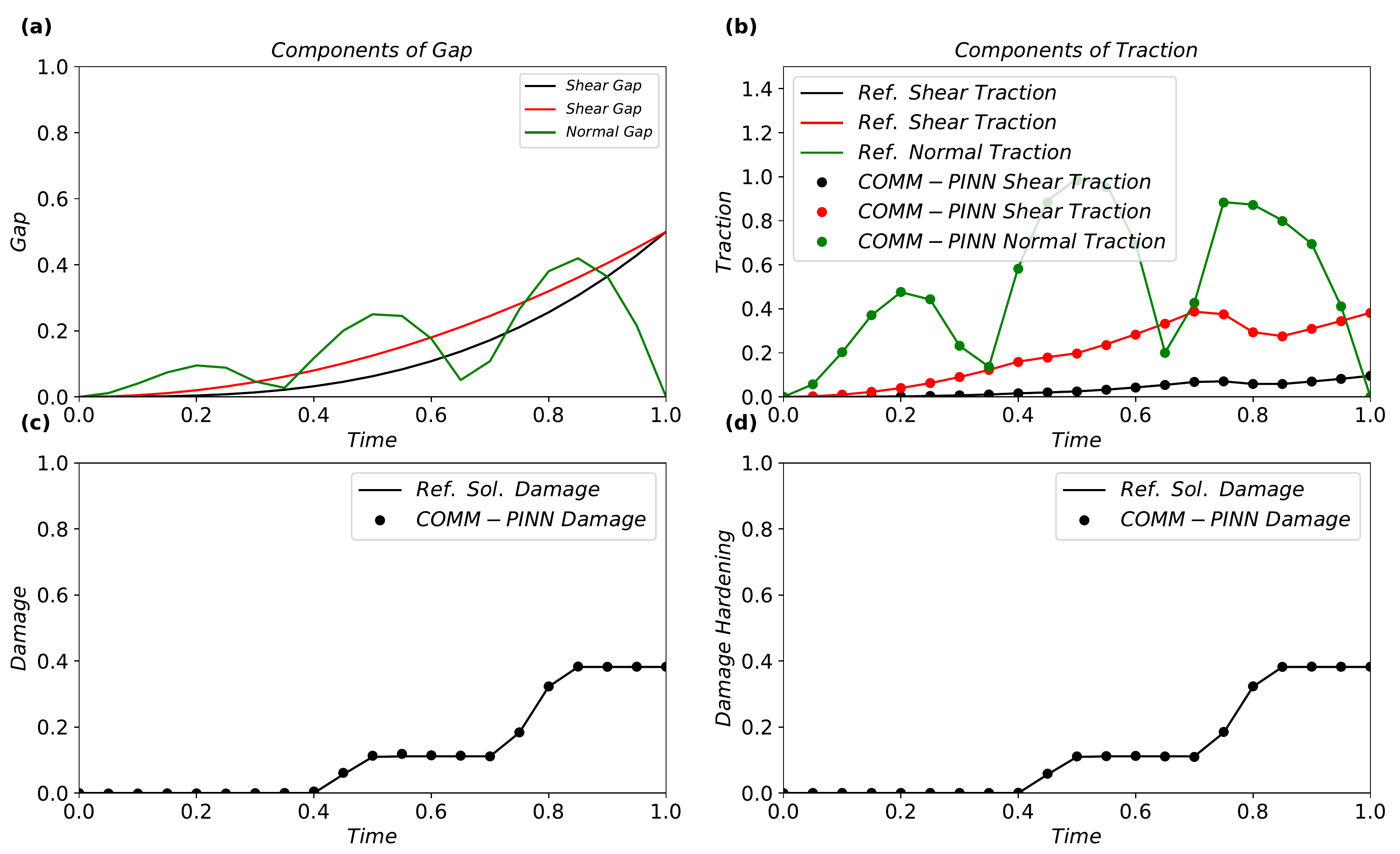}
  \caption{Given gap vector and predicted traction vector for the introduced 3D cohesive zone model. The components of the gap vector are $g_{s1}(t) = 0.5 t^3$ and $g_{s2}(t) = 0.5 t^2$, $g_n(t) = 0.5 |t~\text{sin}(3\pi t)|$.} 
  \label{fig:3D_load1}
\end{figure}
\begin{figure}[H] 
  \centering
  \includegraphics[width=0.99\linewidth]{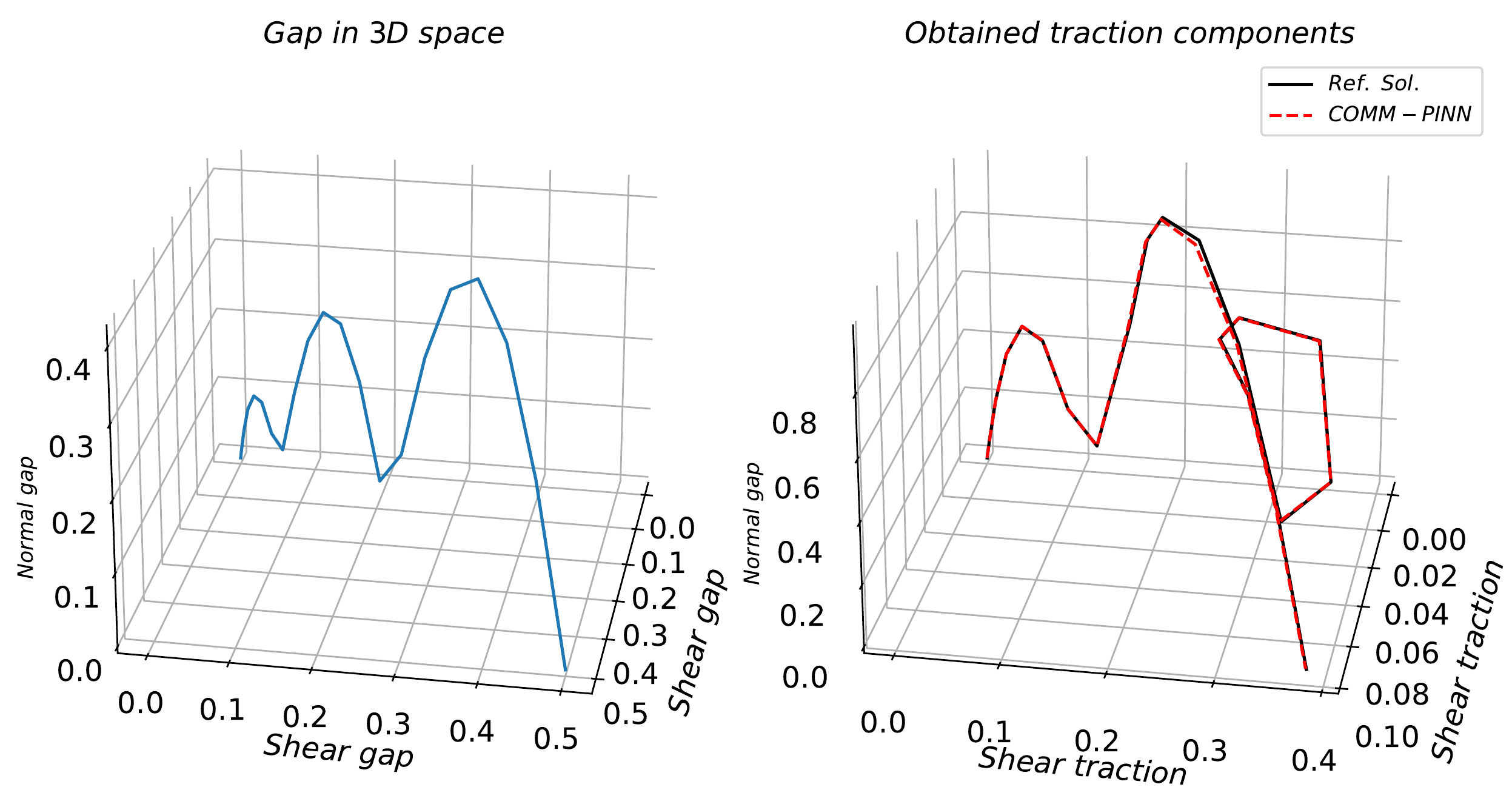}
  \caption{Given gap vector and predicted traction vector for the introduced 3D cohesive zone model.  The components of the gap vector are $g_{s1}(t) = 0.5 t^3$ and $g_{s2}(t) = 0.5 t^2$, $g_n(t) = 0.5 |t~\text{sin}(3\pi t)|$.}
  \label{fig:3D_load1_path}
\end{figure}
\begin{figure}[H] 
  \centering
  \includegraphics[width=0.99\linewidth]{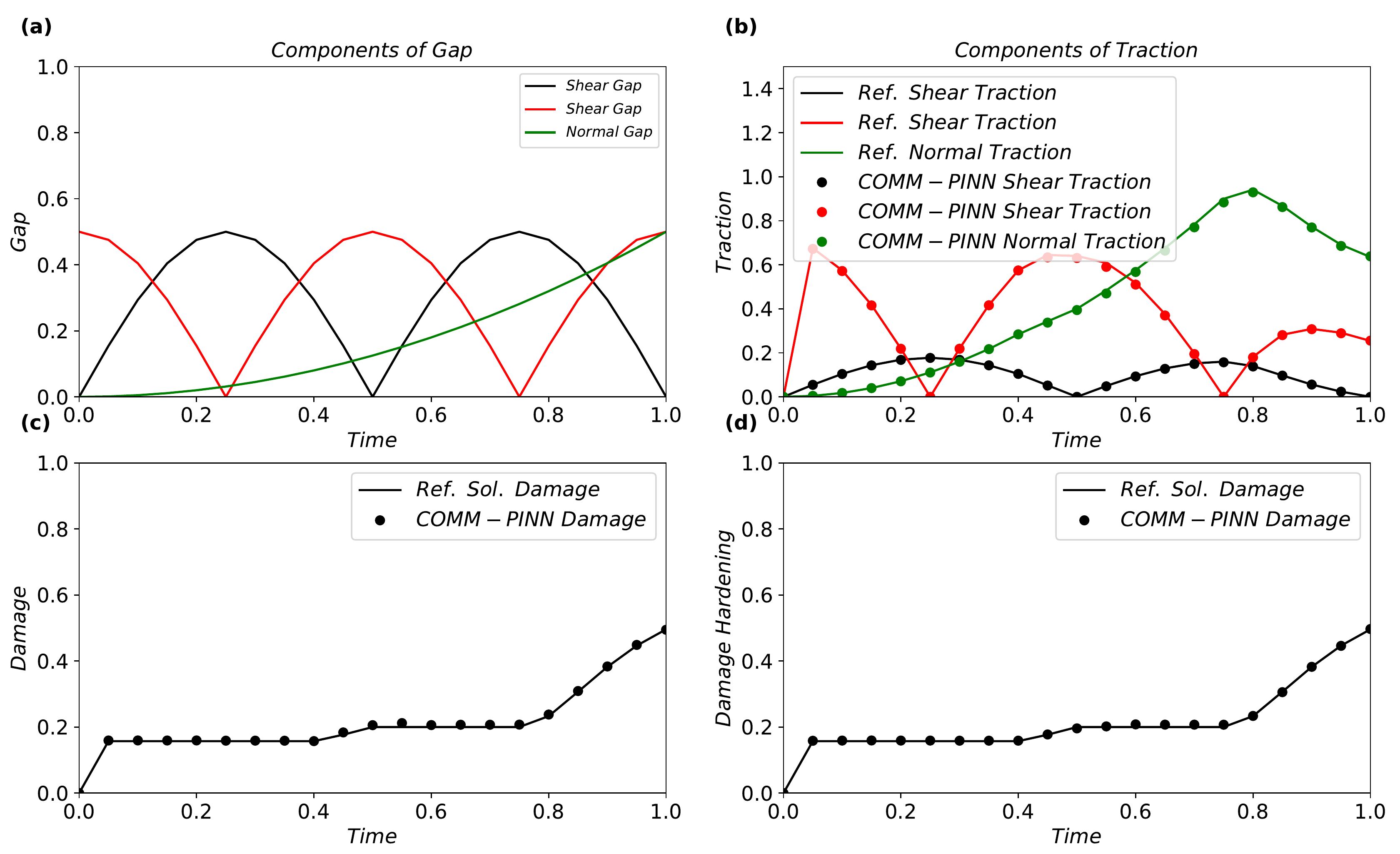}
  \caption{Given gap vector and predicted traction vector for the introduced 3D cohesive zone model. Components of the gap vector are $g_{s1}=0.5 |\text{sin}(2\pi t)|, g_{s2}=0.5|\text{cos}(2\pi t)|, g_{n}=0.5 t^2$} 
  \label{fig:3D_load2}
\end{figure}
\begin{figure}[H] 
  \centering
  \includegraphics[width=0.99\linewidth]{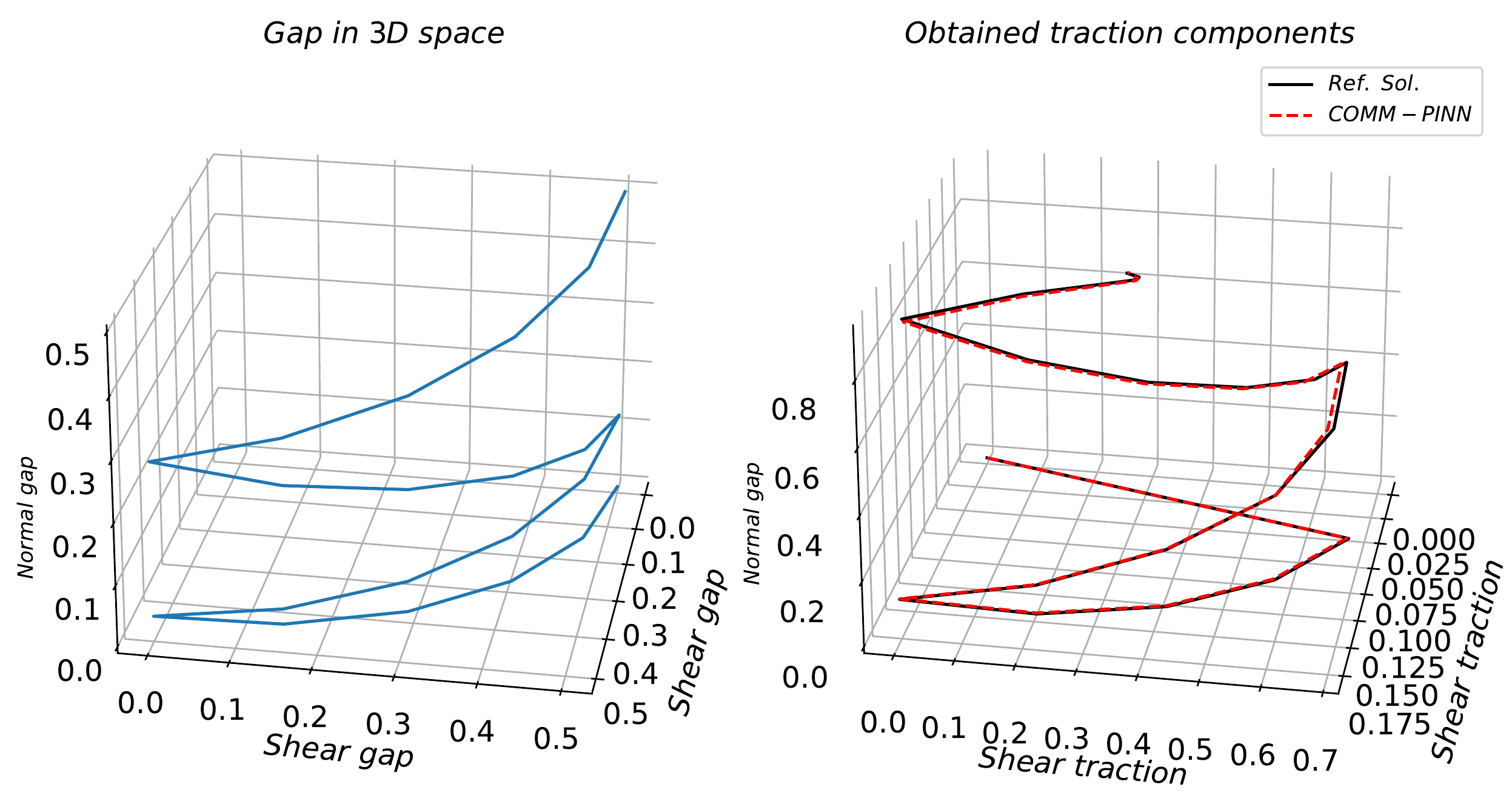}
  \caption{Given gap vector and predicted traction vector for the introduced 3D cohesive zone model. Components of the gap vector are $g_{s1}=0.5 |\text{sin}(2\pi t)|, g_{s2}=0.5|\text{cos}(2\pi t)|, g_{n}=0.5 t^2$.} 
  \label{fig:3D_load2_path}
\end{figure}

In Figures \ref{fig:3D_load1} to \ref{fig:3D_load2_path}, we present the performance of the trained COMM-PINN model compared to the response of the return mapping algorithm. 
In all the results, the first column displays the components of the gap vector, representing the loading direction of the interface in 3D space. The equation corresponding to each loading path is provided in the figure caption, and interested readers can find the relevant codes and data online. The second column presents the obtained components of traction over time. Additionally, the damage value and damage hardening parameter are reported and compared between the two models.

In all the cases, the NN is not exposed to any data from the beginning to the end of the loading process. Only the initial conditions, i.e., the material state, are given. At each time step the NN is capable of predicting the solution. We then use this predicted solution as input to the NN to forecast the subsequent time step. Consequently, any inconsistencies or errors in the predictions from the first step have the potential to propagate to the final step. Nevertheless, we consistently observe superb performance throughout the entire loading history.

\newpage
\subsubsection{Discussions on computational cost for the network evaluation}
\color{black}
Here, we analyze the 3D interface damage model and compare the computational cost between the implicit method (return mapping algorithm), the explicit method, and the COMM-PINN method. 
We conducted multiple tests, running different approaches up to 7 times for one loading path. The averaged outcomes are summarized in Table \ref{tab:network_cost}, demonstrating more than 50\% improvement in computational cost for a single loading path compared to implicit methods.
\color{black}

It is worth noting that the return mapping algorithm was implemented in Python, and efforts were made to optimize the code for efficiency. Following training, the weights and biases of the neural network are called and evaluated using a simple Python code based on the standard feed-forward neural network algorithm (see Eq.~\ref{eq:NN_1}). \textcolor{black}{The network architecture, in this case, consists of 3 layers with 40 neurons in each layer as well another NN with 2 layers with 8 neurons in each layer.} Interestingly, once the NN is trained, the evaluation cost is primarily determined by the number of layers and neurons in the architecture. In other words, evaluating the Eq.~\ref{eq:NN_1} is the sole task we must perform for any arbitrary given loading path. See also arguments and discussions in \cite{EGHBALIAN2023105472}. 

Consequently, when employing a similar architecture, the evaluation time for the NN is nearly identical for both the 1D and 3D cases. Indeed, for 3D cases, increasing the number of neurons per layer may be necessary to provide the neural network (NN) with greater flexibility. However, based on our investigations, we have found that even with a similar number of neurons and layers, the NN was able to accurately capture the interface behavior in both 1D and 3D scenarios. We confirm that the reported results can also be achieved with the same level of accuracy by reducing the number of layers from 5 to 3 or 2. It is important to note that this observation may not universally apply to other types of material models, and further investigation would be required on a case-by-case basis. Therefore, as the dimensionality increases, the advantages of COMM-PINN become increasingly apparent. 

Note that, no additional advanced techniques, such as dropout, were employed to enhance network efficiency. This suggests that further optimization of the neural network architecture should be explored in future developments to reduce the computational cost of network evaluations.

Finally, it is important to mention that the advantages of COMM-PINN become increasingly apparent when dealing with more complex and coupled systems of equations, where return mapping algorithms encounter convergence challenges. For instance, highly coupled (anisotropic) elasto-plastic damage models for ductile fracture pose significant difficulties. However, it is essential to emphasize that exploring such scenarios falls within the scope of future research, and the current work should be regarded as an initial step toward achieving that goal. 

\begin{table}[H]
\centering
\caption{ \textcolor{black}{Comparison of the computational costs between the network evaluation and the return mapping algorithm for the 3D interface damage model.}  }  
\label{tab:network_cost}
\begin{footnotesize}
\begin{tabular}{ l l }
\hline
Method   &  normalized averaged run-time  \\  & to evaluate a random loading path    \\
\hline
Explicit method (see Appendix A)        &   $1.0$  \\
Implicit method / Return mapping algorithm (see Alg.~\ref{alg:dmg} and \cite{REZAEI2019325})        &   $3.4$  \\
COMM-PINN (3 layers with 40 neurons in each)                      &   $1.8$  \\ 
COMM-PINN (2 layers with 10 neurons in each)                      &   $1.2$  \\ 
\hline
\end{tabular}
\end{footnotesize}
\end{table}

\color{black}

\newpage
\subsection{Case studies on the plasticity model}
The relevant network parameters for the case of elastoplasticity are summarized in Table\,\ref{tab:network_pl}.  
\begin{table}[H]
\centering
\caption{Summary of the COMM-PINN network parameters for the plasticity damage model.}  
\label{tab:network_pl}
\begin{footnotesize}
\begin{tabular}{ l l }
\hline
Parameter                          &  Value    \\
\hline
Input, Output                  &   $\{\varepsilon^{i+1}, \varepsilon^{i}_p,\varepsilon^{i}_p\}$, $\{\varepsilon^{i+1}_p,\varepsilon^{i+1}_p\}$ \\ 
Activation function                &  Relu \\ 
Number of layers and neurons per layer ($L$, $N_l$)  &  (5, 80)  \\
Batch size                         &  100  \\ 
Learning rate $\alpha$, number of epochs  &  $(10^{-4},10^{3})$ \\ 
\hline
\end{tabular}
\end{footnotesize}
\end{table} 

There are 4 input parameters: $E$, $\sigma_{y0}$, $h_1$, and $h_2$ for the plasticity model. The chosen values for the elastoplastic model are summarized in Table \ref{tab:plas_parameters}. The current approach may also be used for reverse analysis and calibrating these material parameters based on given data \cite{HAGHIGHAT2023105828}.
\begin{table}[H]
\caption{Material parameters for the plasticity model described in section \ref{sec:plas}.}
	\label{tab:plas_parameters}
	\centering
	\begin{tabular}{ l l l } \hline
	\multirow{1}{*}{}                      & Unit                      & Value                \\ \hline \hline 
	Elastic stiffness $E$       & [MPa]              & $3.0$      \\ 
	Yield stress $\sigma_{\text{y0}}$  & [MPa]                     & $0.6$     \\
	Plasticity hardening parameter $h_1$     & [MPa]                       & $0.4$                \\
	Plasticity hardening parameter $h_2$     & [$-$]                & $10.0$  
    \\ \hline
	\end{tabular}
\end{table}

The location of collocation points is according to the explanations in the previous section. Before we report the results, we introduce two modifications to enhance the NN model performance. The first modification is about balancing the value of different loss functions. For the second modification, we tend to smoothen the hard switch conditions in the plastic model. 

Initially, we used equal weightings for all the loss terms which resulted in an acceptable performance within the training range but does not perform very well when we test the NN beyond the range of the collocation points. After some numerical studies, we concluded that this issue might be due to unbalance nature of different loss terms in the training process. We performed a systematic parameter study and changed the weighting of different loss terms to make them closer to each other. The process is as follows. At first, we start with equal weightings, and then we raise the weighting of those loss functions which are very low in magnitude. By doing so, we observed a huge improvement in the obtained results. This idea is also according to adaptive weighting strategies available in some programming packages. Other adaptive weight techniques should be investigated further in future developments to find more easy-to-use strategies. One of the final optimum results is achieved by selecting $w_{uep}=100$, $w_{uxp}=100$, $w_{evp}=1$, $w_{ylp}=10$, $w_{kep}=100$, $w_{kyp}=10$. 

\begin{figure}[H] 
  \centering
  \includegraphics[width=0.8\linewidth]{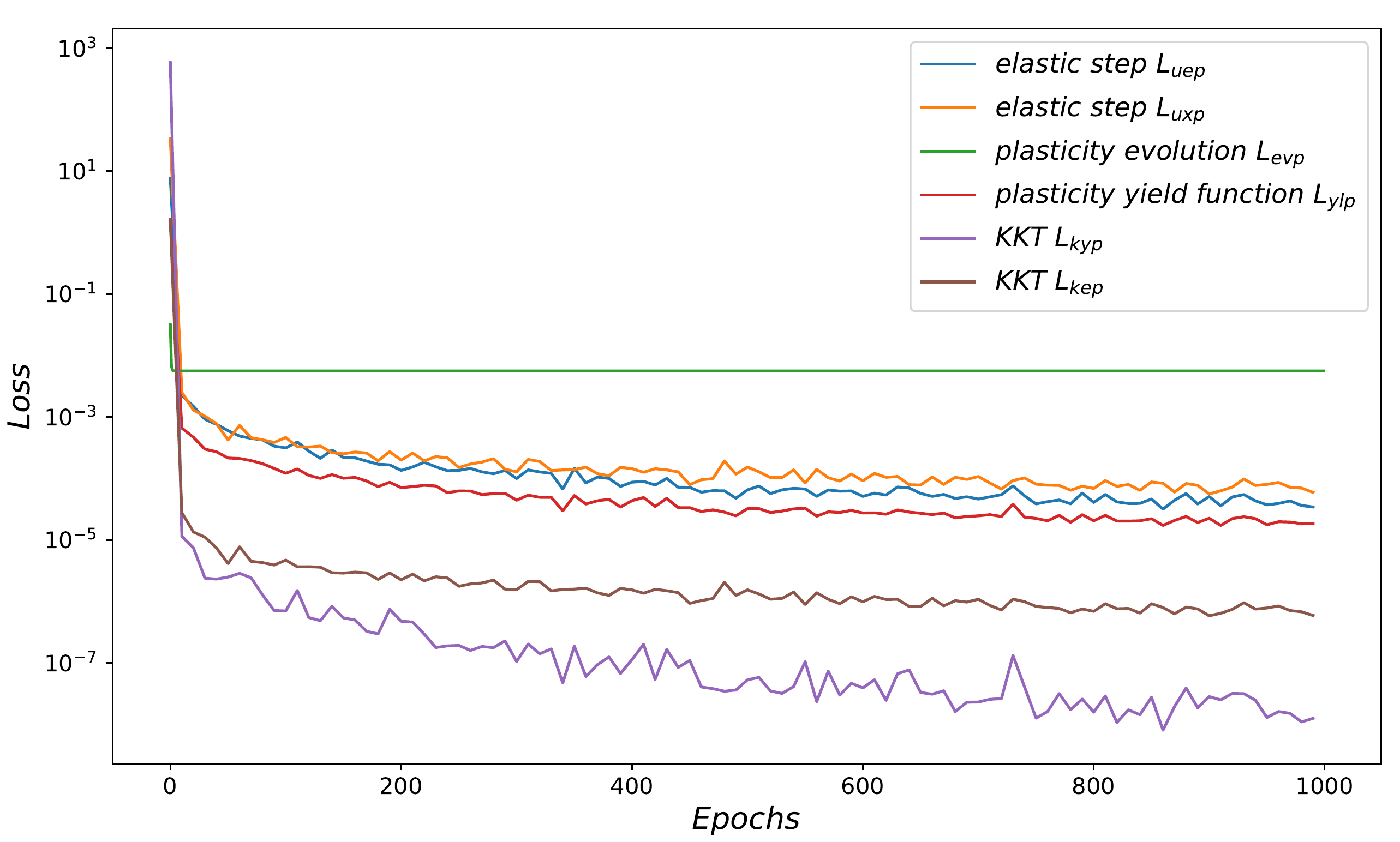}
  \caption{Prominent loss terms for the plasticity model utilizing the proposed COMM-PINN algorithm. }
  \label{fig:loss_plas}
\end{figure}

In the next modification, we loosen the condition of \text{Relu} and \text{sgn} function in Eqs.~\ref{loss_uep} to \ref{loss_kyp}. Therefore, we replace $\text{Relu}(x)$ with $\text{Swish}(x) =x~ \text{Sigmoid}(R x)$, where $R$ determines how sharp is the transition zone. Finally, we replace the $\text{sgn}(x)$ with $ \text{Sigmoid}(R x)$. In the current study, we used $R=300$. To make the influence of the introduced modifications clear to the readers, we perform studies without these modifications as well.
The evolution of each loss term is shown in Fig.~\ref{fig:loss_plas}, where we use $5$ hidden layers with $80$ neurons in each. Furthermore, for the results provided in Fig.~\ref{fig:loss_plas}, we considered the above modifications. The rest of the parameters are according to Table~\ref{tab:network_pl}. Most of the loss functions decay simultaneously, and we did not notice any significant improvement after $1000$ epochs. 

\subsubsection{Influence of weightings in the loss function}
\begin{figure}[H] 
  \centering
  \includegraphics[width=0.99\linewidth]{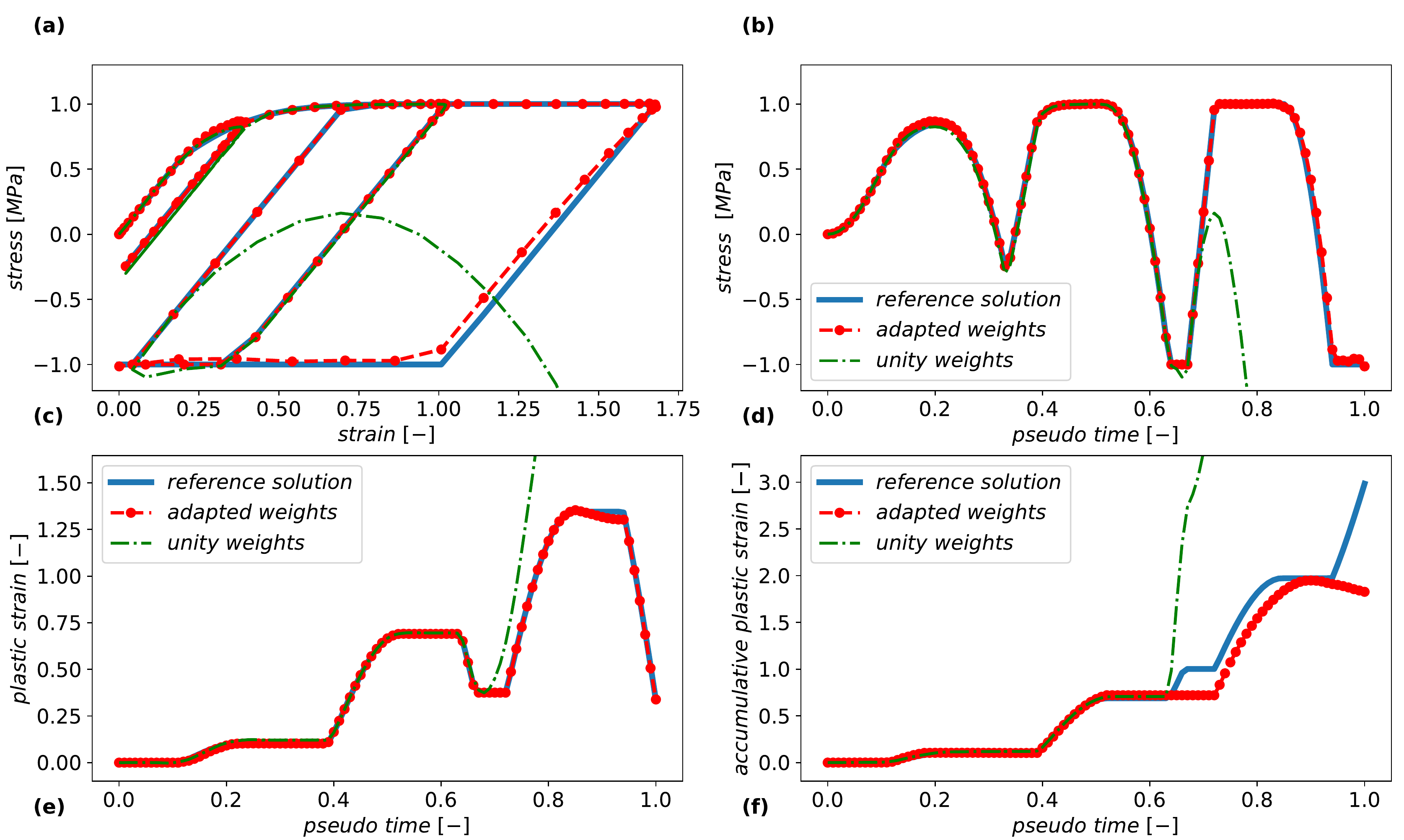}
  \caption{Influence of the adapted weightings for the different loss functions. (a) stress versus strain curve. (b) predicted stress $\sigma(t)$. (c) predicted plastic strain $\epsilon_p(t)$. (d) predicted accumulative plastic strain $\xi_p(t)$. }
  \label{fig:weights_plasticity}
\end{figure}

The studies presented in Fig.\ref{fig:weights_plasticity} were conducted using a loading path of $\varepsilon(t) = 2.0 |t\text{sin}(3\pi t)|$. The time step size used to generate this loading path was $0.01$. The material was initially at rest without any prior history. The results show that the yield point, nonlinear hardening path, and stress saturation for large plastic strains were accurately captured by the NN model. One should note that the NN's response for accumulative plastic strain larger than one can be considered as extrapolation. For the case with equal weights, the performance of the NN is not reliable beyond the range of the collocation points. However, we observed that by tuning and balancing the loss functions, the NN was able to accurately capture the material behavior even beyond the range of the collocation points.

\subsubsection{Modifying switch option in the loss functions}
\begin{figure}[H] 
  \centering
  \includegraphics[width=0.99\linewidth]{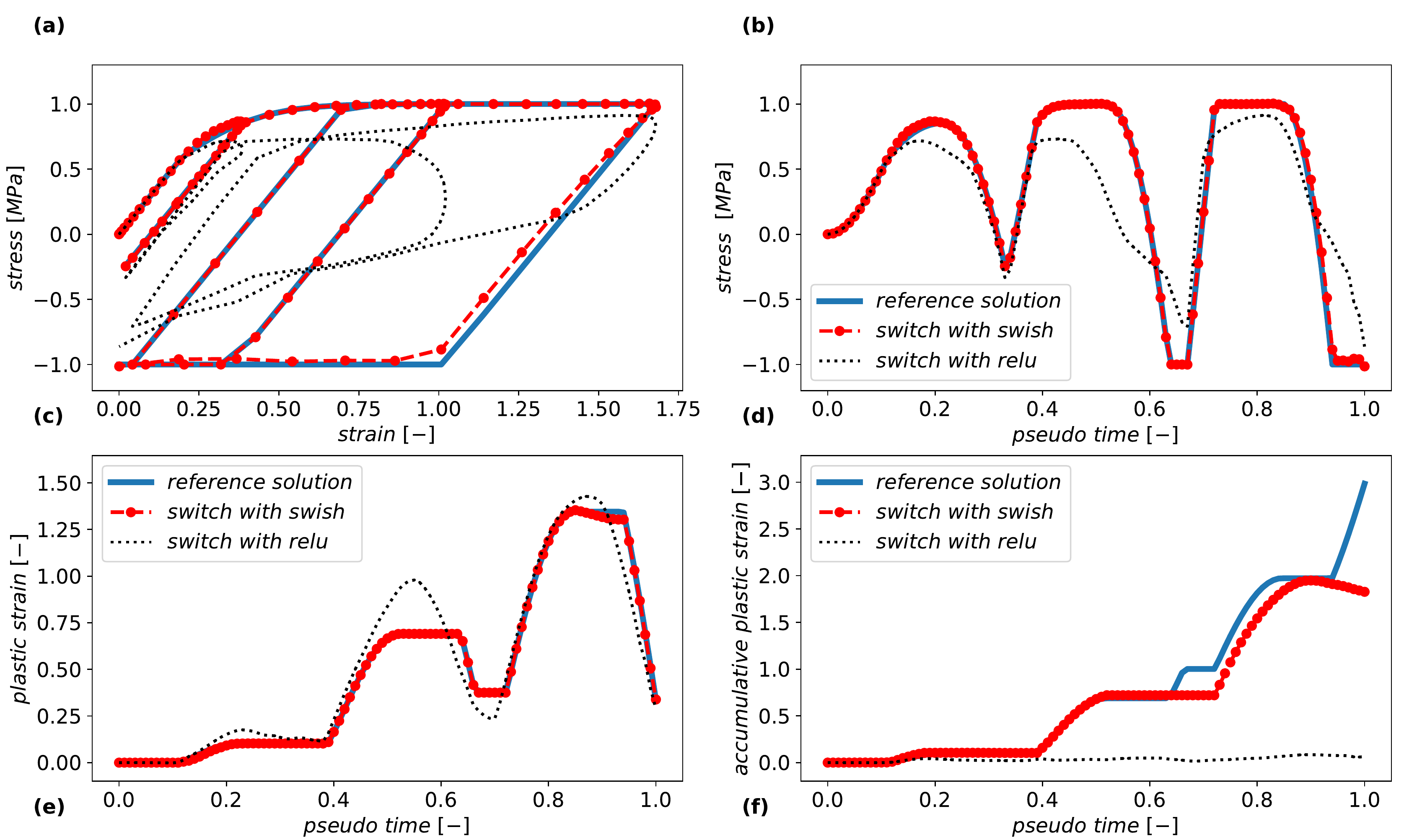}
  \caption{Effect of modifying the switch option in the loss functions. (a) stress versus strain curve. (b) predicted stress $\sigma(t)$. (c) predicted plastic strain $\epsilon_p(t)$. (d) predicted accumulative plastic strain $\xi_p(t)$.}
  \label{fig:form_plasticity}
\end{figure}

In Fig.\ref{fig:form_plasticity}, we studied the case where the function Relu is used for switching between different elastic and plastic conditions (see also Eqs.\ref{loss_uep} to \ref{loss_kyp}). We found that using a smoother version of this function (i.e., $\text{Relu}(x)$ with $x~ \text{Sigmoid}(R x)$) is more beneficial. Our findings indicate that this is particularly important for the condition in $\text{sgn}(x)$, which is replaced by $ \text{Sigmoid}(R x)$, where we use $R=300$. See also investigations by \citet{HAGHIGHAT2023105828}.

\subsubsection{Influence of the time step size}
\begin{figure}[H] 
  \centering
\includegraphics[width=0.99\linewidth]{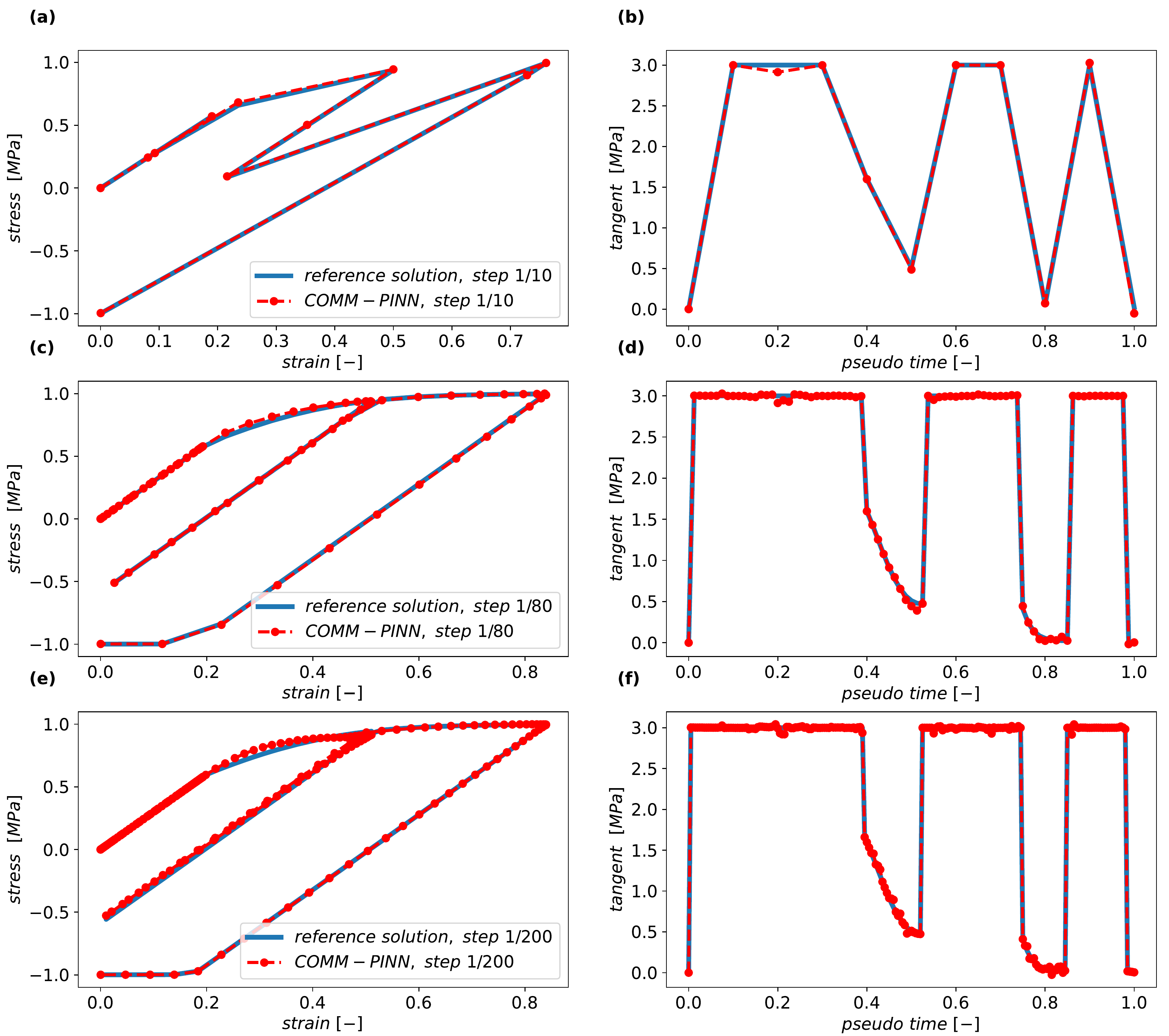}
  \caption{Obtained stress-strain curves and the tangent operator via the train NN and classical material model routine. The time step size for creating the loading path is refined to evaluate the performance of the NN for a strain path that does not fully overlap with the collocation points. }
  \label{fig:tangent}
\end{figure}

Here, we examine the output by resolving the loading path with different time step sizes. These time steps are chosen in a way to ensure that they do not overlap with the collocation points. In the first row of Fig.~\ref{fig:tangent}, the NN is evaluated using a time step size of $0.1$. In addition to the stress-strain curve, we also plot the tangent operator, which is computed solely via automatic differentiation. For the second row, we evaluate the model using a step size of $1/80$ to ensure that there is no overlap with the collocation points (which are generated using a step size of $1/100$). Again, the predictions are acceptable. However, as we use finer time steps (see third row with a step size of $1/200$), we observe some deviation from the correct solution, and the tangent operator shows slight oscillations. As a remedy, one can always utilize finer steps in the generation of the collocation points to improve the accuracy of the NN predictions.

\subsubsection{Comparison with the pure data-driven approach}
\begin{figure}[H] 
  \centering
  \includegraphics[width=0.99\linewidth]{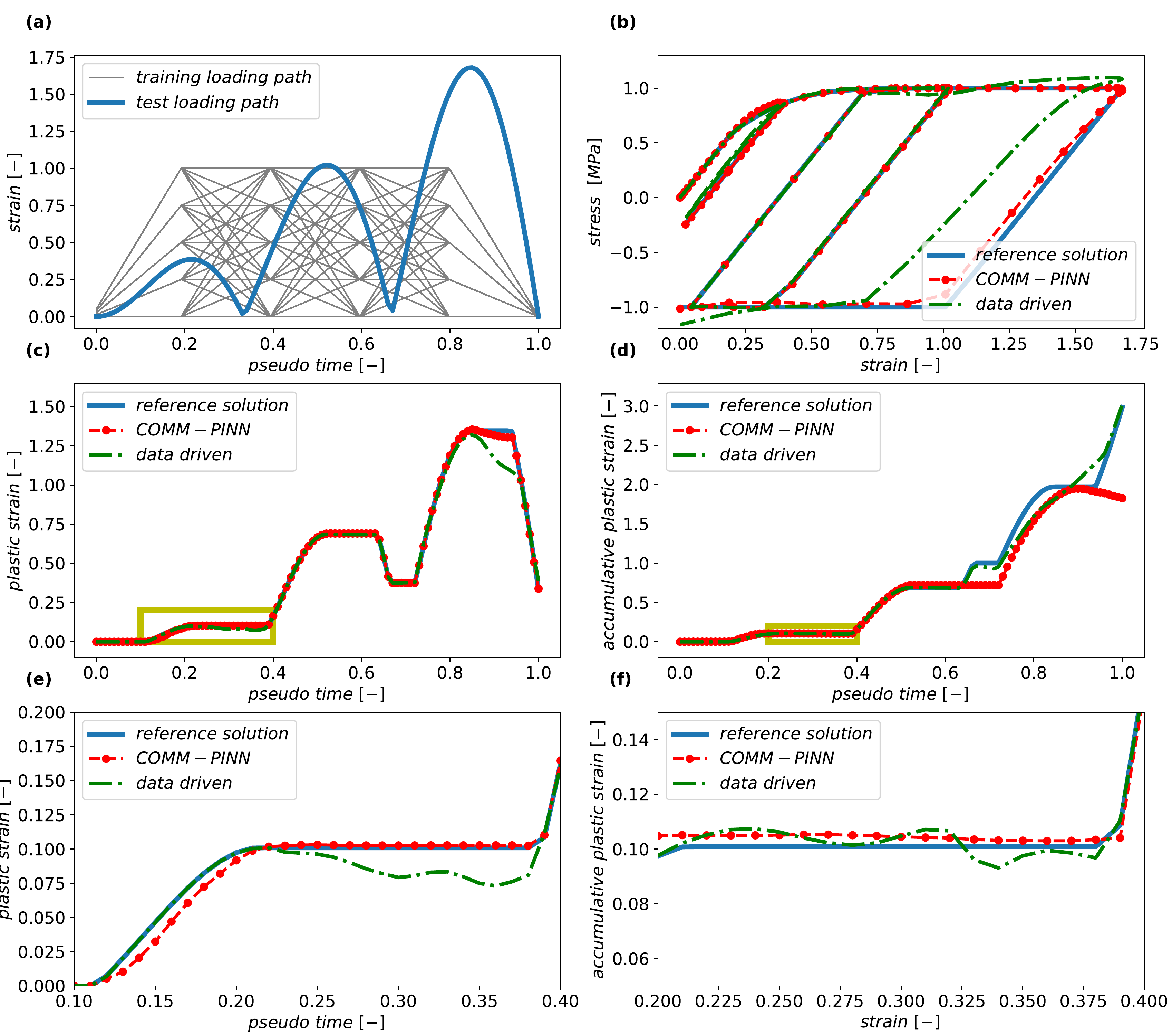}
  \caption{Comparison of the proposed approach against pure data-driven methods. (a) applied strain path is shown in blue and all the loading paths used for training the data-driven method are shown in gray. (b) stress versus strain curve. (c) predicted plastic strain $\epsilon_p(t)$. (d) predicted accumulative plastic strain $\xi_p(t)$.
  }
  \label{fig:data_driven}
\end{figure}

Remark 3 suggests that a purely data-driven approach can be used for the same analysis. In this approach, the data on the incremental evolution of the state variables is obtained from a solution of the governing equations on some random loading paths using the return mapping algorithm (see Alg.\ref{alg:plas}). Once the data is obtained, we use the same neural network architecture described in Table \ref{tab:network_pl}. In this case, the loss function is based on the difference between the predicted and actual values of plastic and accumulative plastic strain, i.e., $\mathcal{L}_{data} = \text{MSE}\left(\varepsilon^{i+1}_p-\varepsilon^{i+1}_{alg}\right) + \text{MSE}\left(\xi^{i+1}_p-\xi^{i+1}_{alg}\right)$. We generated $625$ loading paths for data generation, which are shown in gray in part (a) of Fig.\ref{fig:data_driven}. All loading paths are limited to a strain of 1, which is similar to the range of collocation points. Furthermore, each loading path is discretized by a time step size of $0.01$, which gives us a total of $62500$ data points. Although the initial data points and collocation points are similar, it is not straightforward to have an identical initial condition for these two neural networks.

In Fig. \ref{fig:data_driven}, we compare the two methodologies and their ability to predict material behavior for complex unseen loading paths. However, in parts (e) and (f) of Fig. \ref{fig:data_driven}, we examine the response of these methods in more detail during the first unloading cycle. While the data-driven method shows acceptable predictions, it violates some fundamental physical aspects, such as the decay of plastic strain during elastic unloading, which is unacceptable. Conversely, the results of the COMM-PINN method follow the expected straight line.
For the final loading cycle, we evaluate the extrapolation capabilities of the NN models from both the physics-based and purely data-driven methods. Both methods may fail when extrapolating to extreme conditions. However, the physics-based method with physical constraints performs slightly better. In future developments, it would be interesting to explore the combination of these methods. By adding more collocation points outside the training region, we can enhance the prediction of the hybrid NN model, particularly in cases where there are limited data points for a limited strain interval.

\subsubsection{Influence of collocation points}
\begin{figure}[H] 
  \centering
  \includegraphics[width=0.99\linewidth]{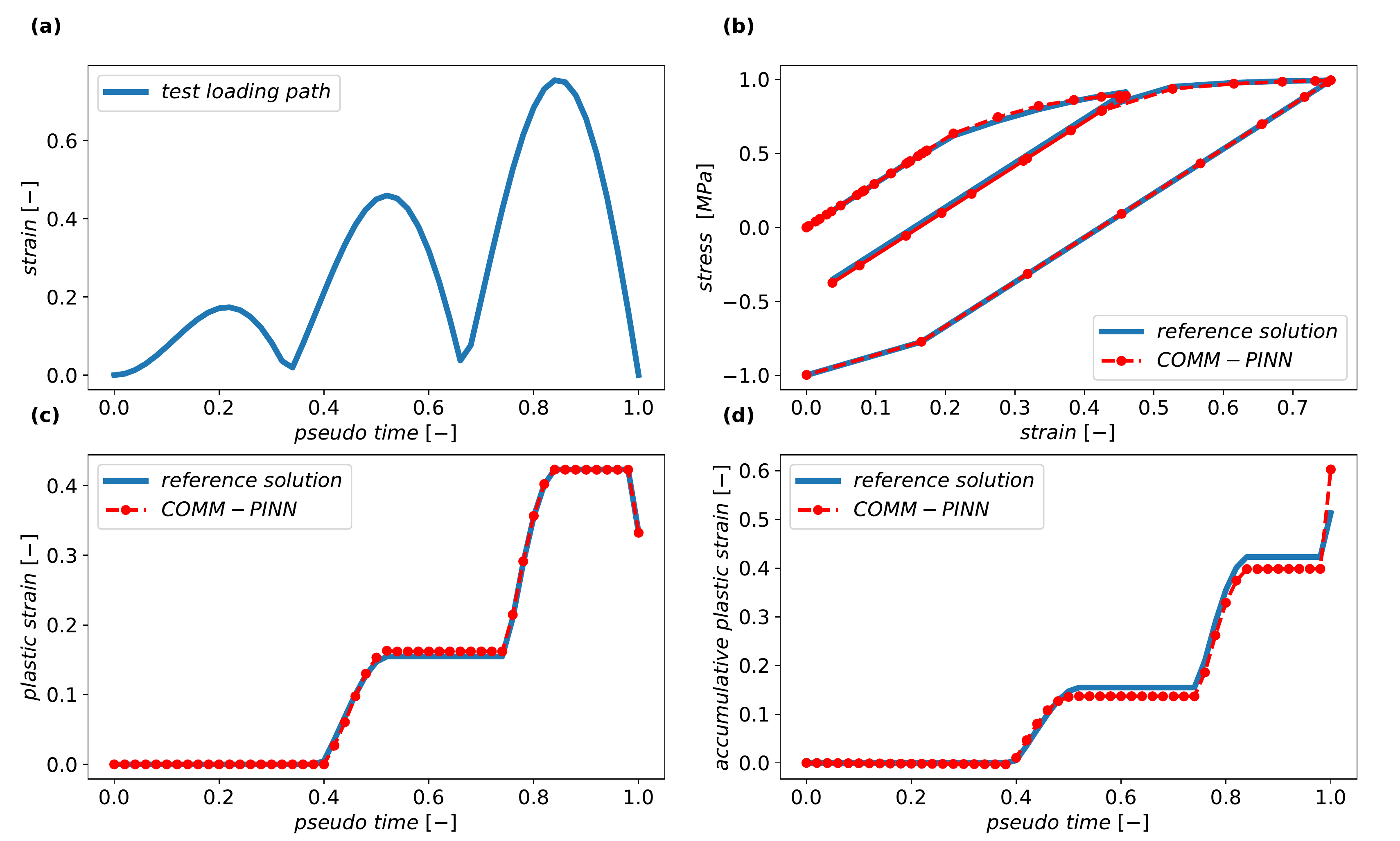}
  \caption{The trained COMM-PINN demonstrates accurate predictions in the evolution of state variables within the range of the collocation points.}
  \label{fig:commpinn_3}
\end{figure}

In the preceding section, it is evident that when we exceed the range of collocation points, the accuracy of the results may decrease due to the violation of the loss functions in this region. However, it is important to note that this should not be perceived as a weakness of the approach, as the range can always be expanded based on the specific application. To provide further clarity on this matter, we conducted two additional tests. In Fig.~\ref{fig:commpinn_3}, we narrowed down the loading range, and it is apparent that the achieved results, particularly for the accumulative plastic strain, displayed significant improvement.

\begin{figure}[H] 
  \centering
  \includegraphics[width=0.99\linewidth]{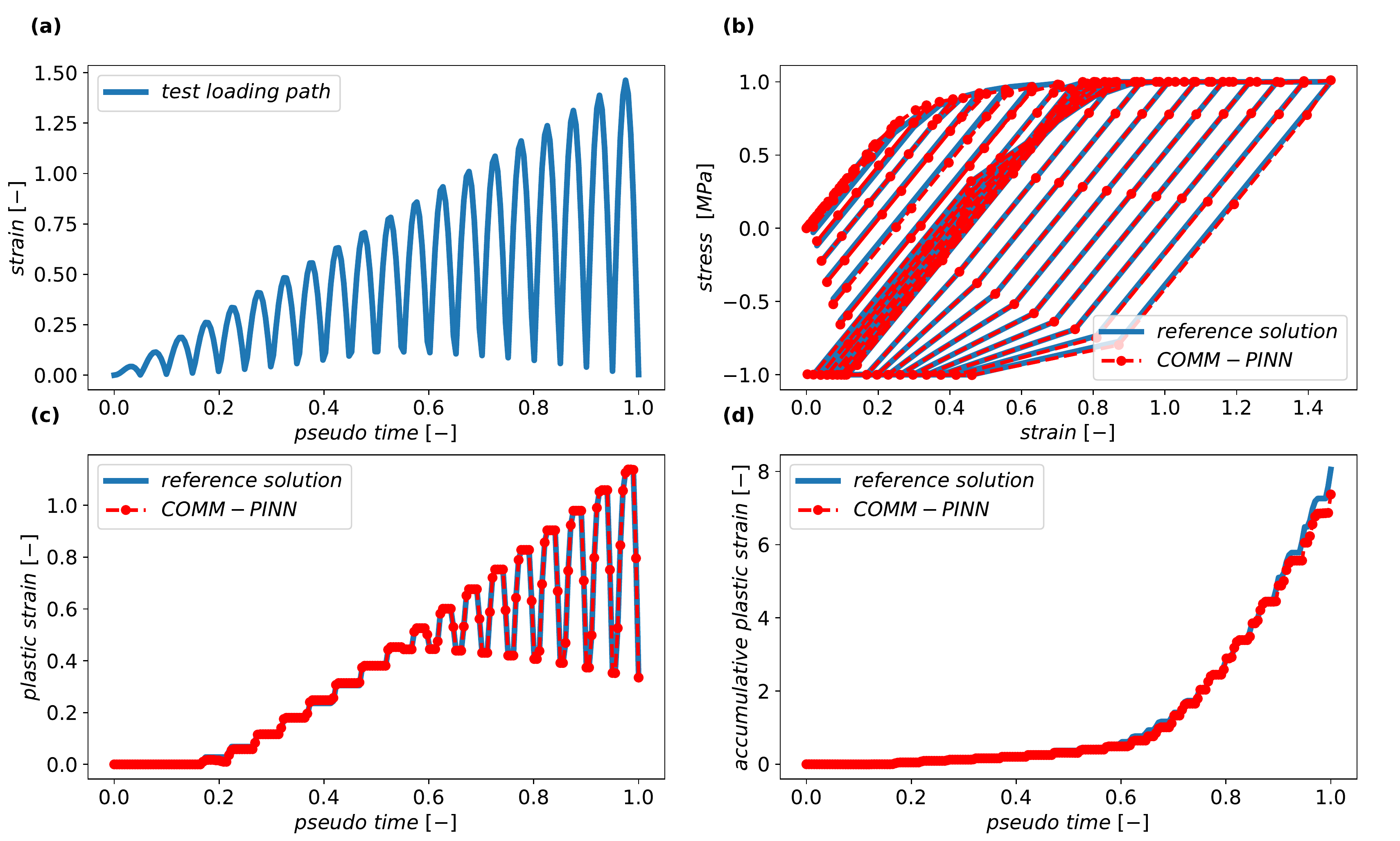}
  \caption{By expanding the range of the collocation points, the trained COMM-PINN exhibits accurate predictions for more complex loading scenarios.}
  \label{fig:commpinn_20}
\end{figure}

Next, we attempted to increase the range of collocation points to 4 for a given strain, plastic strain, and accumulative plastic strain (i.e., $en_{\xi_p} = en_{\varepsilon_p} = en_e = 4.0$). For more details, refer to algorithm \label{alg:data_gen_plas} in section 5. The other training parameters remain the same as before and are explained in the preceding sections. The results obtained under extreme loading conditions are depicted in Fig.~\ref{fig:commpinn_20}. Interestingly, we observed an almost perfect alignment between the two approaches, even when subjected to severe cyclic loading. Note that the chosen loading path is a complex function, where initially the unloading does not return completely to zero but eventually, it converges back to the origin. This selection is intentional, as it allows us to evaluate the performance of the trained neural network under challenging and complicated loading conditions. Further optimization of the network parameters can potentially enhance the results and reduce the margin of error.

\color{black}

\subsubsection{Integration of the COMM-PINN model with finite element method}
\begin{figure}[H] 
  \centering
  \includegraphics[width=0.99\linewidth]{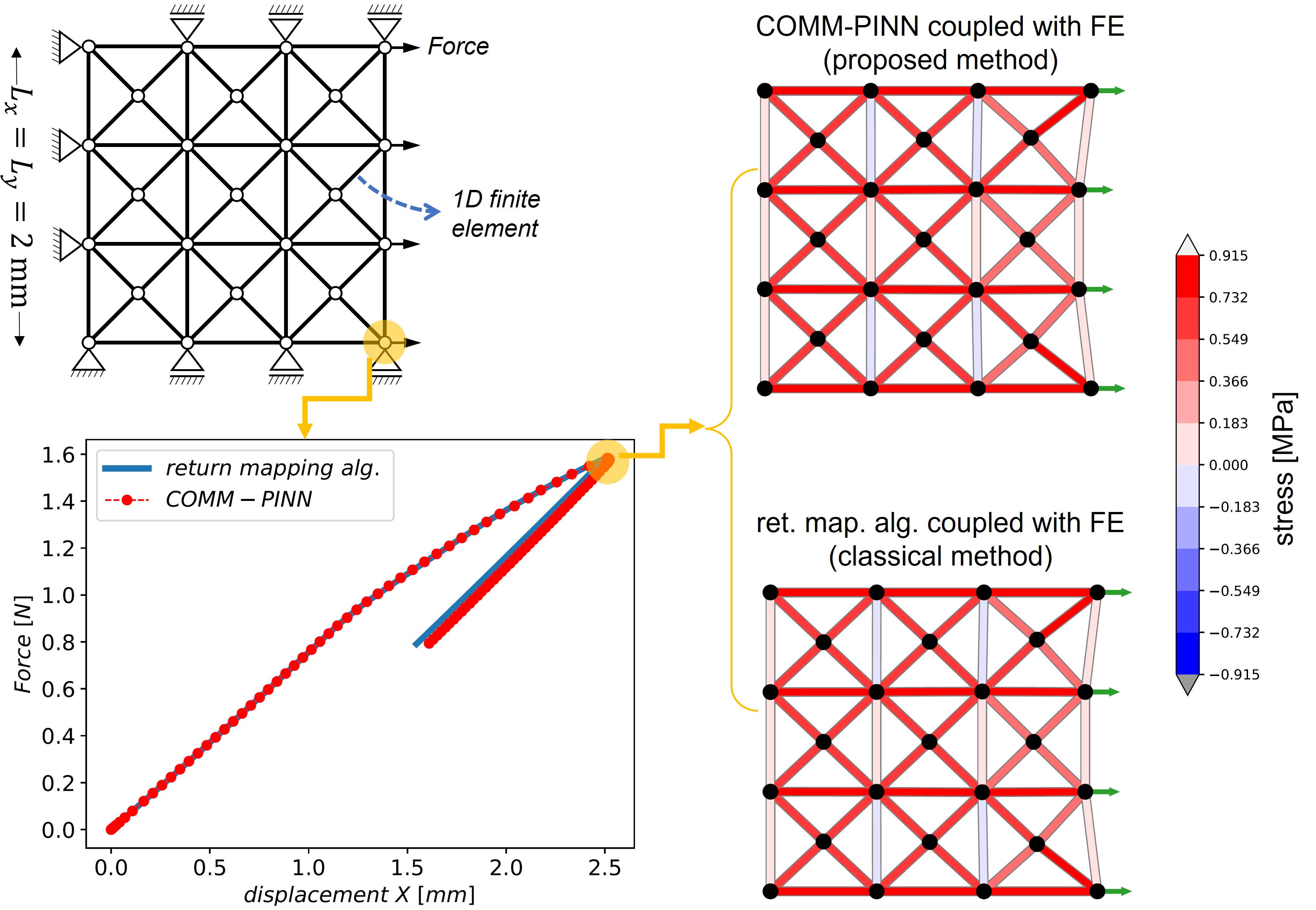}
  \caption{Comparison of the proposed approach against classical return mapping algorithm when they are integrated into the finite element package (trusspy \cite{Dutzler_trusspy_Truss_Solver}). Top left: the geometry and boundary conditions for a simple truss-based structure. Bottom: obtained reaction force from two methodologies. Right: distribution of stress in the truss-based system, obtained from the two methodologies.
  }
  \label{fig:FE}
\end{figure}

We tested the performance of the COMM-PINN model by integrating it into a finite element package. Since the model is trained based on a 1D analysis, we examined a truss-based structure (or a simplified meta-material) where each rod is represented by a single 1D finite element mesh using standard linear shape functions. For this purpose, we utilized the package \textit{trusspy} \cite{Dutzler_trusspy_Truss_Solver}. The geometry and applied boundary conditions are shown in the upper part of Fig.~\ref{fig:FE}, where we fixed the whole structure on the left-hand side and loaded it from the right vertical edge. After the loading phase, the whole system was unloaded.

The obtained reaction force from this process is shown in the lower part of Fig.\ref{fig:FE}. We first performed finite element calculations based on the standard return mapping algorithm and introduced a Voce-type plasticity model. The results are shown in blue. The same boundary value problem was calculated using the trained COMM-PINN model instead of the standard user-defined material model. By comparing the red and blue curves, we conclude that the new method has the potential to replace the classical method in a finite element calculation without any changes to the available codes. Finally, on the right-hand side of Fig.\ref{fig:FE}, we compared the stress distribution from the two methodologies at the end of the loading path and observed identical results.

´
\section{Conclusion and outlooks}
This work demonstrates the potential of physics-informed neural networks in solving constitutive relations for material mechanics. The approach is applicable when the free energy function and internal variables of the model are known beforehand. We proposed strategies to construct the loss functions  for training the network, utilizing expressions for thermodynamic forces, evolution laws, and yield functions. This eliminates the need for labeled data during training. We also suggested techniques to simplify the process and reduce differentiation degrees for evaluating the tangent operator.

Compared to classical return mapping algorithms and other purely data-driven methods, the proposed COMM-PINN method offers several advantages. The model can instantly predict the material response for any loading path within the range of the collocation points. Furthermore, after successful training, the user has the flexibility to freely change the timestep size while maintaining reliable predictions. This capability demonstrates the robustness of the methodology in adapting to different temporal discretizations and its interpolation capabilities. By tuning the weighting of loss functions and performing hyperparameter studies, the model's extrapolation capabilities can be enhanced. The proposed method avoids complex linearizations and yields the tangent operator naturally from the neural network. Labeled data is not necessary for model training, and everything is based on the governing material equations derived from standard thermodynamic principles. Interestingly, the proposed approach showed better performance than purely data-driven methods for a complex test loading path. Finally, the trained material routine can be easily integrated into any finite element package.

The proposed methodology offers a promising opportunity to accelerate the complex material modeling process while also presenting room for further improvement. Our preliminary studies on the 3D damage model for the interface have demonstrated a noteworthy reduction of approximately 50\% in computational cost when compared to standard return mapping algorithms and implicit numerical schemes. This enhancement allows for more efficient simulations of representative volume elements, enabling easier access to full-field simulation results obtained through finite element computations.

By adopting this methodology, users can now primarily focus on developing the material model itself, as the numerical evaluation process, regardless of its complexity, is significantly simplified. Additionally, models implemented in this manner can be effortlessly transferred between different research groups and integrated into various numerical solvers. This ease of integration is attributed to the standardized and cost-effective nature of the network evaluation process, which is achieved through the optimization of the number of neurons and layers.

\color{black}
Under the same time step size, which results in a stable solution for the explicit algorithm (i.e., for $\Delta t=0.0005$), the explicit method for the 3D interface damage model is found to be 3 to 4 times faster than the implicit method and requires almost the same computational cost as the COMM-PINN approach. However, it is important to note that for this specific example and material model, we can increase the time step size by 100 times for the implicit and COMM-PINN approaches and still obtain consistent results.
This indicates that when computing a loading path we can increase the step size to $\Delta t=0.05$. Now, the implicit approach is approximately $100/3.4\sim30$ times faster than the explicit method, while the COMM-PINN approach is about $100/1.2\sim80$ times faster than the explicit method. Therefore, one can conclude that the COMM-PINN approach enjoys the benefits of each method.
\color{black}

In future studies, it will be necessary to extend the framework and train the system for other material properties. Investigations into multi-dimensional cases \cite{ZHANG2020102732} and more complex material routines with anisotropy \cite{IBRAGIMOVA2021103059} are also required to better understand the advantages of this approach. As further examples, one can mention advanced non-local material models as well as developments in multiphysics environment \cite{REZAEI2021104612}. Furthermore, the set of collocation points is needed to be optimally designed to reduce the training time and yet obtain the best accuracy \cite{Shoghi2022}. Ideally, the NN can be trained once for each material model and used as a substitute for standard material subroutines, eliminating the need to rewrite codes and material models in different programming languages.
\\ \\
\textbf{Data Availability}:\\
The codes are available at \href{https://github.com/phyml4e/PINNs/tree/main/COMM_PINN}{COMM-PINN}.

\newpage
\color{black}
\section{Appendix A: comparisons with the explicit method} 
In our study, we prioritized the development of a 3D interface damage model and successfully implemented the explicit method. This is achieved by evaluating the damage parameter using the damage yield function, which relies on information from the previous time step. As a result, it becomes possible to avoid solving additional equations, greatly simplifying the implementation aspects. However, it is important to note that we observed certain issues with the explicit method's stability. To achieve reliable results, we found that one needs to use very fine time-step sizes, such as $\Delta t=0.0005$. In contrast, the implicit method demonstrated greater stability and allowed for reliable predictions even with larger time-step sizes, such as $\Delta t=0.05$, as shown in the figure below. Same holds for the COMM-PINN approach which can be used for much bigger time step sizes (see Fig.~\ref{fig:tangent}).
\begin{figure}[H] 
  \centering
  \includegraphics[width=0.99\linewidth]{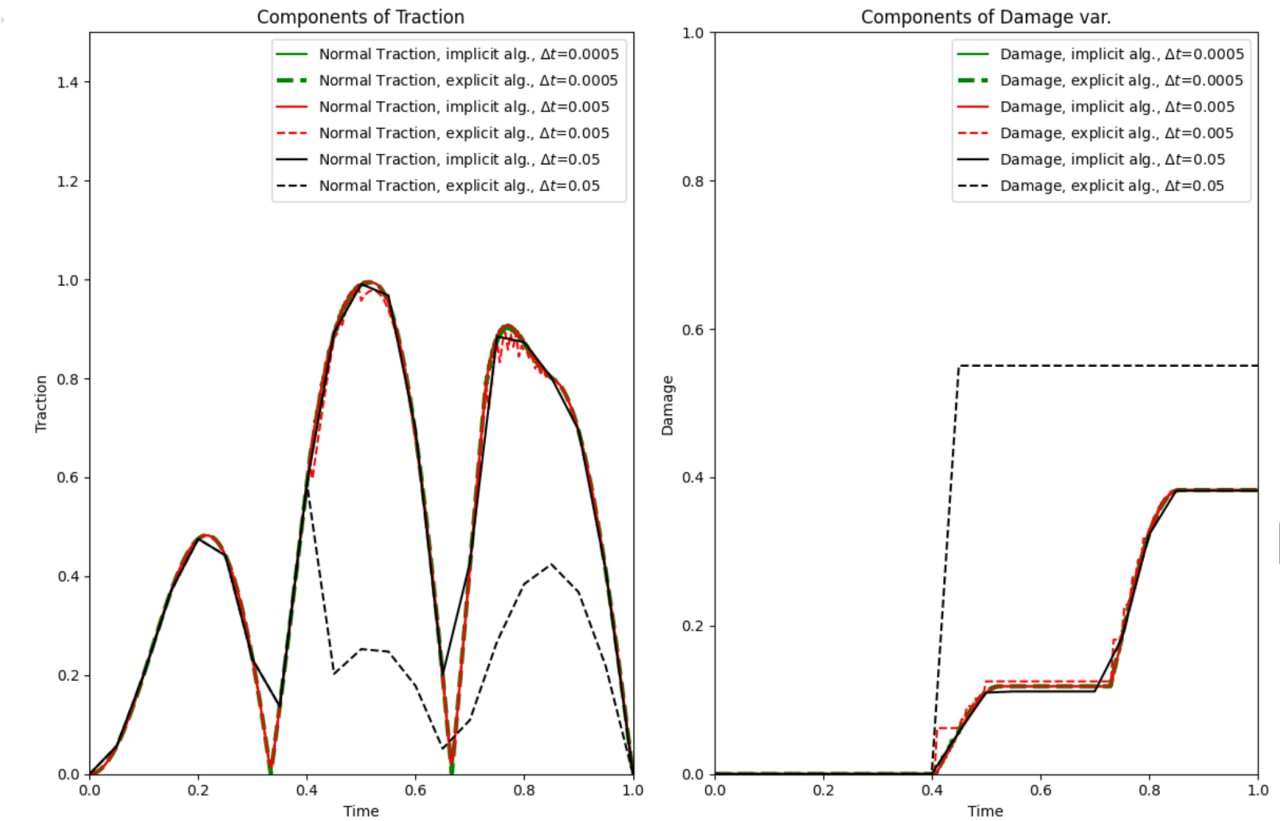}
  \caption{Comparison of the results from different methods and different time step size.} 
  \label{fig:append}
\end{figure}
By employing the implicit method (or COMM-PINN), we can take advantage of larger time steps. These methods not only excel in terms of computational cost but also offer greater reliability. With the implicit methods, we can be confident that the thermodynamic consistency is satisfied, as the underlying equations are solved directly. On the contrary, when using the explicit method, we heavily rely on the results from the previous time step, without assurance that the governing equations are accurately satisfied at the current time. This can lead to oscillatory responses, as observed in the traction behavior for $\Delta t=0.005$.
\color{black}
\\ \\
\textbf{Author Statement}:\\
S.R.: Conceptualization, Software, Supervision, Writing - Review \& Editing. A. M.: Software, Writing - Review \& Editing. A. H.: Software, Writing - Review \& Editing.

\bibliography{Ref}

\end{document}